\documentclass[journal,a4paper]{IEEEtran}
\usepackage[T1]{fontenc}
\usepackage{graphicx,booktabs,multirow,bm}
\usepackage{algorithm}
\usepackage{algpseudocode}
\usepackage{array}
\usepackage{tabularx}
\usepackage{amssymb}
\usepackage{amsmath} 
\usepackage{xcolor}
\ifCLASSINFOpdf
\else
\fi
\usepackage{amsmath}
\usepackage[cmintegrals]{newtxmath}
\ifCLASSOPTIONcompsoc
  \usepackage[caption=false,font=normalsize,labelfont=sf,textfont=sf]{subfig}
\else
  \usepackage[caption=false,font=footnotesize]{subfig}
\fi
\begin{document}

%
\title{Learning Task-Oriented Communication for Edge Inference: An Information Bottleneck Approach}
%
%
%

\author{Jiawei~Shao,~\IEEEmembership{Student Member,~IEEE,}
        Yuyi~Mao,~\IEEEmembership{Member,~IEEE,}
        and~Jun~Zhang,~\IEEEmembership{Senior Member,~IEEE}
\thanks{J. Shao and J. Zhang are with the Department of Electronic and Computer Engineering, The Hong Kong University of Science and Technology, Hong Kong (E-mail: jiawei.shao@connect.ust.hk, eejzhang@ust.hk). Y. Mao is with the Department of Electronic and Information Engineering, Hong Kong Polytechnic University, Hong Kong (E-mail: yuyi-eie.mao@polyu.edu.hk). (The corresponding author is J. Zhang.)}
}

\maketitle

\begin{abstract}
This paper investigates task-oriented communication for edge inference, where a low-end edge device transmits the extracted feature vector of a local data sample to a powerful edge server for processing. 
It is critical to encode the data into an \emph{informative} and \emph{compact} representation for low-latency inference given the limited bandwidth.
We propose a learning-based communication scheme that jointly optimizes feature extraction, source coding, and channel coding in a task-oriented manner, i.e., targeting the downstream inference task rather than data reconstruction.
Specifically, we leverage an information bottleneck (IB) framework to formalize a rate-distortion tradeoff between the informativeness of the encoded feature and the inference performance.
As the IB optimization is computationally prohibitive for the high-dimensional data, we adopt a variational approximation, namely the variational information bottleneck (VIB), to build a tractable upper bound.
To reduce the communication overhead, we leverage a sparsity-inducing distribution as the variational prior for the VIB framework to sparsify the encoded feature vector.
Furthermore, considering dynamic channel conditions in practical communication systems, we propose a variable-length feature encoding scheme based on dynamic neural networks to adaptively adjust the activated dimensions of the encoded feature to different channel conditions.
Extensive experiments evidence that the proposed task-oriented communication system achieves a better rate-distortion tradeoff than baseline methods and significantly reduces the feature transmission latency in dynamic channel conditions. 
\end{abstract}


%
\IEEEpeerreviewmaketitle

%
%
%
%

\section{Introduction}

The recent revival of artificial intelligence (AI) has led to their adaptations in a broad spectrum of application domains, ranging from speech recognition \cite{speech_graves2013speech} and natural language processing (NLP) \cite{natureLP}, to computer vision \cite{CV_survey} and augmented/virtual reality (AR/VR) \cite{vr}. 
Most recently, the potential of AI technologies has also been exemplified in communication systems \cite{chen2019artificial,downey2020machine}. Aiming at delivering data with extreme levels of reliability and efficiency, various design problems of \emph{data-oriented communication}, including transceiver structures \cite{o2017introduction}, source/channel coding \cite{jssc_deniz}, signal detection \cite{samuel2019learning}, and radio resource management \cite{shen2020graph}, have been revisited intensively using AI techniques, especially deep neural networks (DNNs), breeding the emerging area of ``\emph{learning to communicate}''.
It is widely perceived that learning-driven techniques are critical complements to traditional model-driven approaches for communication system designs that rely heavily on expert knowledge, and will undoubtedly transform the wireless networks toward the next generation \cite{letaief2019roadmap}.

Meanwhile, emerging AI applications also raise new communication problems \cite{zhu2020toward,shi2020communication}. To provide an immersive user experience, DNN-based mobile applications need to be performed within the edge of wireless networks, which eliminates the excessive latency incurred by routing data to the Cloud, and is referred to as \emph{edge inference} \cite{edge_intelligence}, \cite{shi2020communication}. 
Edge inference can be implemented by deploying DNNs at an edge server located in close proximity to mobile devices, known as \emph{edge-only inference}.
However, the transmission latency remains a bottleneck for applications with stringent delay requirements \cite{vr,maqueda2018event_driving,liu2019edge_ar}, as a huge volume of data (e.g., 3D images, high-definition videos, and point cloud data) need to be uploaded. 
On the other hand, the resource-demanding nature of DNNs often makes it infeasible to be deployed as a whole locally for \emph{device-only inference} due to the limited on-device computational resources \cite{cai2019once}.

\emph{Device-edge co-inference} appears to be a prominent solution for fast edge inference \cite{edge_intelligence,kang2017neurosurgeon,JALAD}, which reduces the communication overhead by harvesting the available computational resources at both the edge servers and mobile devices.
A mobile device first extracts a compact feature vector from the raw input data using an affordable neural network and then uploads it for server-based processing.
Nevertheless, most existing device-edge co-inference proposals simply split a pre-trained DNN into two subnetworks to be deployed at a device and a server, leaving feature compression and transmission to a traditional communication module \cite{JALAD}.
Such kind of decoupled treatment ignores the interplay between wireless communications and the inference tasks, and thus fails to exploit the full benefits of collaborative inference since the communication strategies can be adaptive to specific tasks.
To address this limitation and improve the inference performance, in this paper, we propose a \emph{task-oriented communication} principle for edge inference and develop an innovative learning-driven approach under the framework of information bottleneck (IB) \cite{tishby2000informationIB}.

\subsection{Related Works and Motivations}

The line of research on ``learning to communicate'' stems from the introductory article on deep learning for the physical layer design in \cite{o2017introduction}, where information transmission was viewed as a data reconstruction task, and a communication system can thus be modeled by a DNN-based autoencoder with the wireless channel simulated by a non-trainable layer. 
The autoencoder-based framework for communication systems was later extended to a deep joint source-channel coding (JSCC) architecture for wireless image transmission in \cite{jssc_deniz}, which enjoys significant improvement of image reconstruction quality over separate source/channel coding techniques.
JSCC has also been applied to natural language processing for text transmission, which was accomplished by incorporating the semantic information of sentences using recurrent neural networks \cite{Goldsmith}. 
It is worth noting that the aforementioned works focus on \emph{data-oriented communication}, which targets at transmitting data reliably given the limited radio resources.

Nevertheless, the shifted objective of feature transmissions for accurate edge inference with low latency is not aligned with that of data-oriented communication, as it regards a part of the raw input data (e.g., nuisance, task-irrelevant information) as meaningless.
Thus, recovering the original data sample with high fidelity at the edge server results in redundant communication overhead, which leaves room for further compression. 
This insight is also supported by a basic principle from representation learning \cite{bengio2013representation}: A good representation should be insensitive (or invariant) to nuisances such as translations, rotations, occlusions. Thus, we advocate for \emph{task-oriented communication} for applications such as edge inference, to improve the efficiency by transmitting \emph{sufficient} but \emph{minimal} information for the downstream task.

There have been recent studies on feature compression for efficient transmission in edge inference \cite{wen2016learning,pruning,iccshao,jankowski2020wireless_Jankowski,shao2020communication}.
In particular, for the image classification task, an end-to-end architecture was proposed in \cite{iccshao} to jointly optimize the feature compression and encoding by integrating deep JSCC.
In contrast to data-oriented communication that concerns the data recovery metrics (e.g., the $l_{2}$-distance or bit error rate), the proposed method was directly trained with the cross-entropy loss for the targeted classification task and ignored the data reconstruction quality.
The end-to-end training facilitates the mapping of task-relevant information to the channel symbols and omits the irrelevance.
Similar ideas were utilized to design feature compression and encoding schemes for image retrieval tasks at the wireless network edge in \cite{jankowski2020deep} and for point cloud data processing in \cite{shao2020branchy}. 

While the end-to-end learning-driven architectures for task-oriented communication have been proven effective in saving communication bandwidth, there remain multiple restrictions unsolved in order to unleash their highest potentials: First, there lacks a systematic way to quantify the informativeness of the encoded feature vector and its impact on the inference tasks, hindering to achieve the best inference performance given the available resources; Besides, the dynamic wireless channel condition necessitates adaptive encoding scheme for reliable feature transmission, which has received less attention in existing frameworks (e.g. \cite{iccshao,jankowski2020wireless_Jankowski,shao2020communication,NJSCC}). 
These form the main motivations of our study.

Data-oriented communication relies on classical source coding and channel coding theory, which, however, is not optimized for task-oriented communication. Recently, an information theoretical design principle, named information bottleneck (IB) \cite{tishby2000informationIB}, has been applied to investigate deep learning, which seeks the right balance between data fit and generalization by using the mutual information as both a cost function and a regularizer.
Particularly, the IB framework maximizes the mutual information between the latency representation and the label of the data sample to promote high accuracy, while minimizing the mutual information between the representation and the input sample to promote generalization. Such a tradeoff between preserving the \emph{relevant} information and finding a \emph{compact} representation fits nicely with bandwidth-limited edge inference and thus will be adopted as the main design principle in our study for task-oriented communication. 
The IB framework is inherently related to the communication problem of remote source coding (RSC) \cite{RSC}.
It has recently attracted great attention from both the machine learning and information theory communities \cite{goldfeld2020information,zaidi2020information,achille2018information_dropout,alemi2016deepVIB}.
Nevertheless, applying it to task-oriented communication demands additional optimization, which forms the main technical contributions of our study.

\subsection{Contributions}

In this paper, we develop effective methods for task-oriented communication for device-edge co-inference based on the IB principle \cite{tishby2000informationIB}. Our major contributions are summarized as follows:
\begin{itemize}
\item We design the task-oriented communication system by formalizing a rate-distortion tradeoff using the IB framework. Our formulation aims at maximizing the mutual information between the inference result and the encoded feature, meanwhile, minimizing the mutual information between the encoded feature and input data. Thus, it addresses the objectives of improving the inference accuracy, while reducing the communication overhead, respectively. To the best of our knowledge, this is the first time that IB is introduced to design wireless edge inference systems.
\item As the mutual information terms in the IB formulation are generally intractable for DNNs with high-dimensional features, we leverage the variational approximation, known as variational information bottleneck (VIB), to devise a tractable upper bound. Besides, by selecting a sparsity-inducing distribution as the variational prior, the VIB framework identifies and prunes the redundant dimensions of the encoded feature vector to reduce the communication overhead. The proposed method is named as \emph{variational feature encoding} (VFE).
\item We extend the proposed task-oriented communication scheme to dynamic communication environments by enabling flexible adjustment of the transmitted signal length. In particular, we develop a \emph{variable-length variational feature encoding} (VL-VFE) based on dynamic neural networks that can adaptively adjust the active dimensions according to different channel conditions.
\item The effectiveness of the proposed task-oriented communication schemes is validated in both static and dynamic channel conditions on image classification tasks. Extensive simulation results demonstrate that VFE and VL-VFE outperform the traditional communication design and existing learning-based joint source-channel coding for data-oriented communication.
\end{itemize}

\subsection{Organization}

The rest of the paper is organized as follows. Section \ref{system_model} introduces the system model and describes the design objective of task-oriented communication.
Section \ref{basic_method} and Section \ref{variable_length} propose the task-oriented communication schemes in static and dynamic channel conditions, respectively.
In Section \ref{experiment}, we provide extensive simulation results to evaluate the performance of the proposed task-oriented communication schemes. Finally, Section \ref{conclusion} concludes the paper.

\subsection{Notations}

Throughout this paper, upper-case letters (e.g. $X$ and $Y$) and lower-case letters (e.g. $\boldsymbol{x}$ and $\boldsymbol{y}$) stand for random variables and their realizations, respectively. The entropy of $Y$ and the conditional entropy of $Y$ given $X$ are denoted as $H(Y)$ and $H(Y|X)$, respectively. The mutual information between $X$ and $Y$ is represented as $I(X,Y)$, and the Kullback-Leibler (KL) divergence between two probability distributions $p(\bm{x})$ and $q(\bm{x})$ is denoted as $D_{KL}\left(p||q\right)$.
The statistical expectation of $X$ is denoted as $\mathbf{E}\left(X\right)$.
We further denote the Gaussian distribution with mean $\bm{\mu}$ and covariance matrix $\bm{\Sigma}$ as $\mathcal{N}\left(\bm{\mu},\bm{\Sigma}\right)$ and use $\bm{I}$ to represent the identity matrix.

\begin{figure*}[t]
\centering
\subfloat[Data-oriented communication for device-edge co-inference.]{
\label{communication-oriented}
\centering
\includegraphics[width=0.85\linewidth]{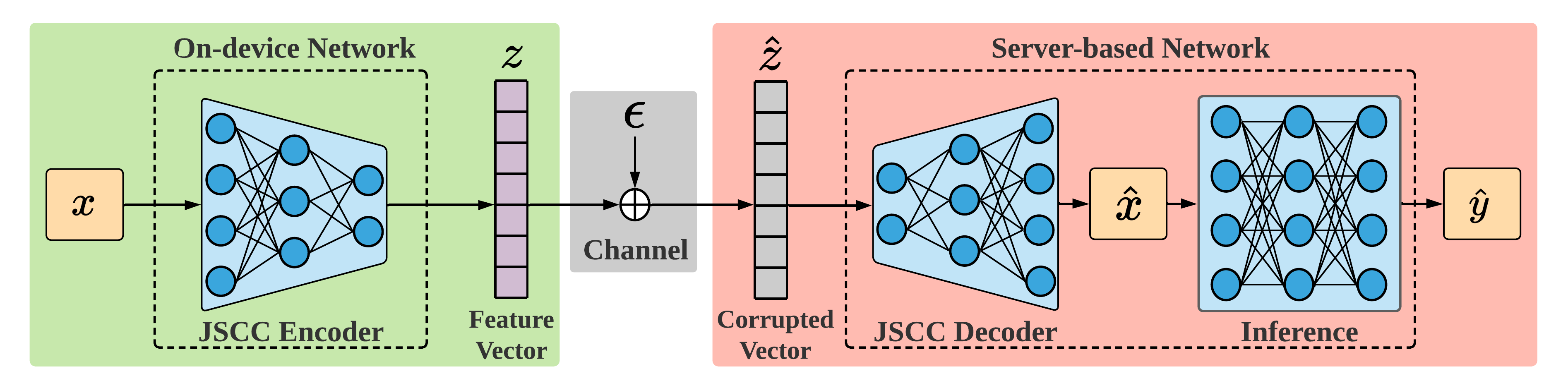}
}
\vfill
\subfloat[Task-oriented communication for device-edge co-inference.]{
\centering
\label{task-orietend}
\includegraphics[width=0.85\linewidth]{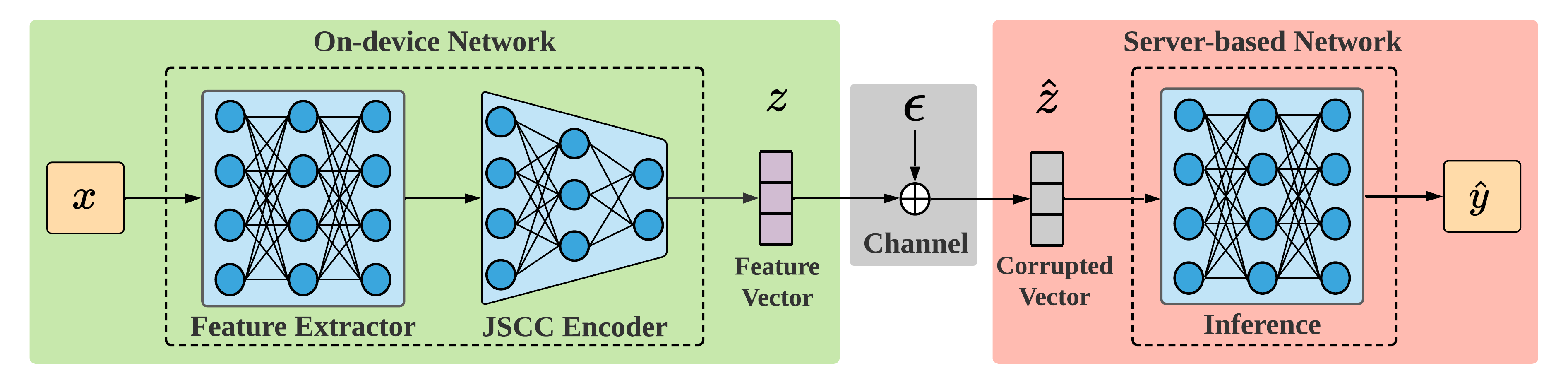}
}

\caption{Two kinds of communication schemes for device-edge co-inference: Learning-based data-oriented and task-oriented communication.
The green region corresponds to a mobile device, and the red region corresponds to an edge server.
In data-oriented communication (top), a mobile device transmits the encoded feature $\boldsymbol{z}$ of the original data $\boldsymbol{x}$ (e.g., an image). Then, an edge server attempts to decode the data $\boldsymbol{\hat{x}}$ based on the noise-corrupted feature $\boldsymbol{\hat{z}}$, and further utilizes $\boldsymbol{\hat{x}}$ as input to obtain the inference result $\boldsymbol{\hat{y}}$ (e.g., the label of input data).
In contrast, task-oriented communication (bottom) extracts and encodes useful information $\boldsymbol{z}$ jointly by the on-device network, and the receiver directly leverages $\boldsymbol{\hat{z}}$ to obtain the inference result $\boldsymbol{\hat{y}}$, without recovering the original data. Therefore, $\boldsymbol{z}$ could be a highly compressed representation since the task-irrelevant information can be discarded.
}
\label{block_diagram}

\end{figure*}

\section{System Model and Problem Description}
\label{system_model}

\subsection{System Model}

We consider task-oriented communication in a device-edge co-inference system as shown in Fig. \ref{task-orietend}, where two DNNs are deployed at the mobile device\footnote{While two components, i.e., a feature extractor and a JSCC encoder, are shown in Fig. \ref{task-orietend} at the device, they can be regarded as a single DNN. We consider resource-constrained devices that can only afford light DNNs, which are unable to complete the inference task with sufficient accuracy. More details of the adopted neural network architecture will be discussed in Section \ref{experiment}.} and the edge server respectively so that they can cooperate to perform inference tasks, e.g., image classification and object detection. 
The input data $\boldsymbol{x}$ and its target variable $\boldsymbol{y}$ (e.g., label) are deemed as different realizations of a pair of random variables $(X,Y)$. The encoded feature, received feature (noise-corrupted feature), and the inference result are respectively instantiated by random variables $Z$, $\hat{Z}$ and $\hat{Y}$. These random variables constitute the following probabilistic graphical model:
\begin{equation}
\label{probability_graphical_model}
    Y \rightarrow X \rightarrow Z \rightarrow \hat{Z} \rightarrow \hat{Y},
\end{equation}
which satisfies $p(\boldsymbol{\hat{y}},\boldsymbol{\hat{z}},\boldsymbol{{z}}|\boldsymbol{x}) = p_{\boldsymbol{\theta}}(\boldsymbol{\hat{y}}|\boldsymbol{\hat{z}})p_{\text{channel}}(\boldsymbol{\hat{z}}|\boldsymbol{z})p_{\boldsymbol{\phi}}(\boldsymbol{z}|\boldsymbol{x})$, with DNN parameters $\boldsymbol{\theta}$ and $\boldsymbol{\phi}$ to be discussed below.

As shown in Fig. \ref{task-orietend}, the on-device network defines the conditional distribution $p_{\boldsymbol{\phi}}(\boldsymbol{z}|\boldsymbol{x})$ parameterized by $\boldsymbol{\phi}$, which consists of a feature extractor and a JSCC encoder.
The extractor first identifies the task-relevant feature from the raw input $\boldsymbol{x}$, and then the JSCC encoder maps the feature values to the channel input symbols $\boldsymbol{z}$.
Since both the extractor and encoder are parameterized by DNNs, these two modules can be jointly trained in an end-to-end manner.
Then, the encoded feature $\boldsymbol{z}$ is transmitted to the server over the noisy wireless channel, and the server receives the noise-corrupted feature $\boldsymbol{\hat{z}}$.
In this paper, we assume a scalar Gaussian channel between the mobile device and the edge server for simplicity, which is modeled as a non-trainable layer with the transfer function denoted as $\boldsymbol{\hat{z}} = \boldsymbol{z} + \boldsymbol{\epsilon}$.
The additive channel noise $\boldsymbol{\epsilon}$ is sampled from a zero-mean Gaussian distribution with $\sigma^{2}$ as the noise variance, i.e., $\boldsymbol{\epsilon} \sim \mathcal{N}(\boldsymbol{0},  {\sigma}^{2}\boldsymbol{I})$.
To account for the limited transmit power at the mobile device, we constrain the power of each dimension of the encoded feature vector to be below $P$, i.e., $z_{i}^2 \leq P, \forall i = 1, \cdots, n$ with $n$ as the encoded feature vector dimension. Thus, the channel condition can be characterized by the peak signal-to-noise ratio (PSNR) defined as follows:
\begin{equation*}
\mathrm{PSNR}=10 \log  \frac{P}{\sigma^{2}} \  (\mathrm{dB}).
\end{equation*}
{Note that although we assume a scalar Gaussian channel model for simplicity, the system can be extended to other channel models as long as we can estimate the channel transfer function \cite{deep_learning_based_ota} and the distribution $p_{\text{channel}}(\bm{\hat{z}}|\bm{z})$.}
Finally, the server-based network leverages $\boldsymbol{\hat{z}}$ for further processing and outputs the inference result $\boldsymbol{\hat{y}}$ with the distribution $p_{\boldsymbol{\theta}}(\boldsymbol{\hat{y}}|\boldsymbol{\hat{z}})$ parameterized by $\boldsymbol{\theta}$.

\subsection{Problem Description}
The communication overhead is characterized by the number of nonzero dimensions of the output of the JSCC encoder. Intuitively, if symbols over more dimensions are transmitted, the edge server will get a high-quality feature vector, which leads to higher inference accuracy, but it will induce a higher communication overhead and latency. So there is an inherent tradeoff between the inference performance and the communication overhead, which is a key ingredient for the design of task-oriented communication. This can be regarded as a new and special kind of \emph{rate-distortion tradeoff}.
Therefore, we resort to the information bottleneck (IB) principle \cite{tishby2000informationIB} to formulate an optimization problem that minimizes the following objective function\footnote{Note that the IB objective function is unrelated to the parameter $\boldsymbol{\theta}$ since the distribution $p(\boldsymbol{y}|\boldsymbol{\hat{z}})$ is defined by $p(\boldsymbol{x},\boldsymbol{y})$, $p_{\boldsymbol{\phi}}(\boldsymbol{z}|\boldsymbol{x})$, and $p_{\text{channel}}(\boldsymbol{\hat{z}}|\boldsymbol{z})$.}:
\begin{align}
    \mathcal{L}_{IB}(\boldsymbol{\phi}) =  & \underbrace{-I(\hat{Z}, Y)}_{\text {Distortion }}+ \beta  \underbrace{I(\hat{Z}, X)}_{\text {Rate }} \nonumber \\
    = & \ \mathbf{E}_{p(\boldsymbol{x},\boldsymbol{y})}\big\{ \mathbf{E}_{p_{\boldsymbol{\phi}}(\boldsymbol{\hat{z}} | \boldsymbol{x})}[-\log p(\boldsymbol{y} | \boldsymbol{\hat{z}})] \nonumber \\
    & + \beta D_{K L}\left(p_{\boldsymbol{\phi}}(\boldsymbol{\hat{z}} | \boldsymbol{x}) \| p(\boldsymbol{\hat{z}})\right)\big\} - H(Y) \nonumber \\
    \equiv &  \ \mathbf{E}_{p(\boldsymbol{x},\boldsymbol{y})}\big\{ \mathbf{E}_{p_{\boldsymbol{\phi}}(\boldsymbol{\hat{z}} | \boldsymbol{x})}[-\log p(\boldsymbol{y} | \boldsymbol{\hat{z}})] \nonumber \\ 
    & + \beta D_{K L}\left(p_{\boldsymbol{\phi}}(\boldsymbol{\hat{z}} | \boldsymbol{x}) \| p(\boldsymbol{\hat{z}})\right)\big\}, \label{IB_1}
\end{align}
where the equivalence in the last row is in the sense of optimization, ignoring the constant term $H(Y)$. 
The objective function is a weighted sum of two mutual information terms with $\beta >0$ controling the tradeoff.
Specifically, the quantity $I(\hat{Z}, X)$ is comprehended as the preserved information in $\hat{Z}$ given $X$ and measured by the minimum description length \cite{cover2012elements_IT} (or rate).
Besides, since the entropy of $Y$, i.e., $H(Y)$, is a constant related to the input data distribution, minimizing the term $-I(\hat{Z},Y)$ is equivalent to minimizing the conditional entropy $H(Y|\hat{Z})$, which characterizes the uncertainty (distortion) of the inference result $Y$ given the received noise-corrupted feature vector $\hat{Z}$.
Thus, the IB principle formalizes a rate-distortion tradeoff for edge inference systems, and minimizes the conditional mutual information $I(X,\hat{Z}|Y)$, which corresponds to the amount of redundant information that needs to be transmitted.
Compared with data-oriented communication, the IB framework retains the task-relevant information and results in $I(\hat{Z}, X)$ that is much smaller than $H(X)$, which reduces the communication overhead.

\subsection{Main Challenges}

The IB framework is promising for task-oriented communication
as it explicitly quantifies the informativeness of the
encoded feature vector and offers a formalization of the rate-distortion tradeoff in edge inference. However, there are three main challenges when applying it to develop practical feature encoding methods, listed as follows.
\begin{itemize}
\item \textbf{Estimation of mutual information}: The computation of mutual information terms for high-dimensional data with unknown distributions is challenging since the empirical estimate for the probability distribution requires the sampling number to increase exponentially with the dimension \cite{wang2019nonparametric}. Therefore, developing a tractable estimator for mutual information is critical to make the problem solvable.
\item \textbf{Effective control of communication overhead}: Minimizing the mutual information between the input data and the feature vector indeed reduces the redundancy about task-irrelevant information. However, there is no direct link between redundancy reduction and feature sparsification, which controls the communication overhead with a JSCC encoder. Thus, to reduce the communication overhead, an effective method is needed to aggregate the nuisance to the expandable dimensions so that the number of symbols to be transmitted is minimized.
\item \textbf{Dynamic channel conditions}: The hostile wireless channel always poses significant challenges for communication systems. Particularly, the channel dynamics have to be accounted for. Dynamically adjusting the encoded feature length based on the DNNs is nontrivial, as the neural network structure is fixed since initialization. Changing the activation of neurons according to the channel conditions calls for other control modules.
\end{itemize}
The following two sections will tackle these challenges, and develop effective methods for task-oriented communications. The effectiveness of the proposed methods will be tested in Section \ref{experiment}.

\section{Variational Feature Encoding}
\label{basic_method}

In this section, we develop a variational information bottleneck (VIB) framework to resolve the difficulty of mutual information computation of the original IB objective in (\ref{IB_1}).
Besides, we show that by selecting a sparsity-inducing distribution as the variational prior, minimizing the mutual information between the raw input data $X$ and the noise-corrupted feature $\hat{Z}$ facilitates the sparsification of $\hat{Z}$ by pruning the task-irrelevant dimensions.
Such an activation pruning scheme, i.e., removing neurons in a DNN, is effective in reducing the overhead of task-oriented communication.
Based on this idea, we name our proposed method as variational feature encoding (VFE).
This section assumes a static channel condition, while dynamic channels will be treated in Section \ref{variable_length}.

\subsection{Variational Information Bottleneck Reformulation}
\label{subsection_vib}

The variational method is a natural way to approximate intractable computations based on some adjustable parameters (e.g., weights in DNNs), and it has been widely applied in machine learning, e.g., the variational autoencoder \cite{kingma2013autovae}.
In the VIB framework, the central idea is to introduce a set of approximating densities to the intractable distribution.

Revisiting the probabilistic graphical model in (\ref{probability_graphical_model}), the distribution $p_{\boldsymbol{\phi}}(\boldsymbol{\hat{z}}|\boldsymbol{x})$ is determined by the on-device DNN and the channel model, i.e., $p_{\boldsymbol{\phi}}(\boldsymbol{\hat{z}}|\boldsymbol{x}) = \int p_{\boldsymbol{\phi}}(\boldsymbol{z}|\boldsymbol{x})p_{\text{channel}}(\boldsymbol{\hat{z}}|\boldsymbol{z};\boldsymbol{\epsilon}) d\bm{z}$.
Particularly, as we adopt a deterministic on-device network, $p_{\boldsymbol{\phi}}(\boldsymbol{z}|\boldsymbol{x})$ can be regarded as a Dirac-delta function.
Then, we have $p_{\boldsymbol{\phi}}(\boldsymbol{\hat{z}}|\boldsymbol{x}) = \mathcal{N}\left(\bm{\hat{z}}|\boldsymbol{z}(\boldsymbol{x};\boldsymbol{\phi}), \sigma^{2}\boldsymbol{I}\right)$, where the deterministic function $\boldsymbol{z}(\boldsymbol{x};\boldsymbol{\phi})$ maps $\boldsymbol{x}$ to $\boldsymbol{z}$ parameterized by $\boldsymbol{\phi}$.
For notational simplicity, we rewrite $p_{\boldsymbol{\phi}}(\boldsymbol{\hat{z}}|\boldsymbol{x}) = \mathcal{N}\left(\boldsymbol{\hat{z}}|\boldsymbol{z}(\boldsymbol{x};\boldsymbol{\phi}), \sigma^{2}\boldsymbol{I}\right)$ as $p_{\boldsymbol{\phi}}(\boldsymbol{\hat{z}}|\boldsymbol{x}) = \mathcal{N}\left(\bm{\hat{z}}|\boldsymbol{z}, \sigma^{2}\boldsymbol{I}\right)$.

With a known distribution $p_{\boldsymbol{\phi}}(\boldsymbol{\hat{z}}|\boldsymbol{x})$ and the joint data distribution $p(\boldsymbol{x},\boldsymbol{y})$, the distributions $p(\boldsymbol{\hat{z}})$ and $p(\boldsymbol{y}|\boldsymbol{\hat{z}})$ are fully characterized by the underlying Markov chain $Y \leftrightarrow X \leftrightarrow \hat{Z} $.
Unfortunately, these two distributions are intractable due to the following high-dimensional integrals:
\begin{equation*}
\begin{aligned}
    p(\boldsymbol{\hat{z}}) = & \int p(\boldsymbol{x})p_{\boldsymbol{\phi}}(\boldsymbol{\hat{z}}|\boldsymbol{x})d\boldsymbol{x}, \\
    p(\boldsymbol{y}|\boldsymbol{\hat{z}}) = & \int \frac{p(\boldsymbol{x},\boldsymbol{y})p_{\boldsymbol{\phi}}(\boldsymbol{\hat{z}}|\boldsymbol{x})}{p(\boldsymbol{\hat{z}})}d\boldsymbol{x}.
\end{aligned}
\end{equation*}
To overcome this issue, we apply two variational distributions $q(\boldsymbol{\hat{z}})$ and $q_{\boldsymbol{\theta}}(\boldsymbol{y}|\boldsymbol{\hat{z}})$ to approximate the true distributions $p(\boldsymbol{\hat{z}})$ and $p(\boldsymbol{y}|\boldsymbol{\hat{z}})$, respectively, where $\boldsymbol{\theta}$ is the parameters of the server-based network shown in Fig. \ref{task-orietend} that computes the inference result $\boldsymbol{\hat{y}}$.
Therefore, we recast the objective function in (\ref{IB_1}) as follows:
\begin{equation}
\label{loss_in_3}
\begin{aligned}
\mathcal{L}_{VIB}(\boldsymbol{\phi},\boldsymbol{\theta}) = & \ \mathbf{E}_{p(\boldsymbol{x},\boldsymbol{y})} \left\{ \mathbf{E}_{p_{\boldsymbol{\phi}}(\boldsymbol{\hat{z}} | \boldsymbol{x})}\left[-\log q_{\boldsymbol{\theta}}(\boldsymbol{y} | \boldsymbol{\hat{z}})\right] \right. \\
& + \beta D_{K L}\left(p_{\boldsymbol{\phi}}(\boldsymbol{\hat{z}} | \boldsymbol{x}) \| q(\boldsymbol{\hat{z}})\right)\big\}.
\end{aligned}  
\end{equation}
The above formulation is termed as the variational information bottleneck (VIB) \cite{alemi2016deepVIB}, which invokes an upper bound on the IB objective function in (\ref{IB_1}).
Details of the derivations are deferred to the Appendix \ref{appendix a}.
By further applying the reparameterization trick \cite{kingma2013autovae} and Monte Carlo sampling, we are able to obtain an unbiased estimate of the gradient and hence optimize the objective using stochastic gradient descent. 
In particular, given a mini-batch of data $\left\{\left(\boldsymbol{x_{i}}, \boldsymbol{y_{i}}\right)\right\}_{i=1}^{M}$ and sampling the channel noise $L$ times for each pair $(\boldsymbol{x_{i}},\boldsymbol{y_{i}})$, we have the following empirical estimation:
\begin{equation}
\label{loss_in_3_estimate}
\begin{aligned}
   \mathcal{L}_{VIB}(\boldsymbol{\phi},\boldsymbol{\theta}) \simeq & \frac{1}{M} \sum_{m=1}^{M}\left\{ \frac{1}{L} \sum_{l=1}^{L} \left[-\log q_{\boldsymbol{\theta}}\left(\boldsymbol{y_{m}} |\boldsymbol{\hat{z}}_{\boldsymbol{m},\boldsymbol{l}}\right)\right] \right. \\
    & +  \beta D_{K L}\left(p_{\boldsymbol{\phi}}\left(\hat{\boldsymbol{z}} | \boldsymbol{x}_{\boldsymbol{m}}\right) \| q(\hat{\boldsymbol{z}})\right)\Bigg\}, 
\end{aligned}
\end{equation}
where $\boldsymbol{\hat{z}}_{\boldsymbol{m},\boldsymbol{l}} = \boldsymbol{z_{m}} + \boldsymbol{\epsilon}_{\boldsymbol{m},\boldsymbol{l}}$ and $\boldsymbol{\epsilon}_{\boldsymbol{m},\boldsymbol{l}} \sim \mathcal{N}\left(\mathbf{0}, \sigma^{2} \boldsymbol{I}\right)$.

In the next subsection, we illustrate that minimizing the VIB objective helps to prune the redundant dimensions in the encoded feature vector, and thus it serves as a suitable and tractable objective for task-oriented communication.

\subsection{Redundancy Reduction and Feature Sparsification}
As we leverage the IB principle instantiated via a variational approximation, minimizing the KL-divergence term $D_{KL}\left(p(\boldsymbol{\hat{z}}|x)\|q(\boldsymbol{\hat{z}})\right)$ shall reduce the redundancy in feature $\hat{Z}$.
However, it does not guarantee sparse activations in the feature encoding process.
For example, if the reduced redundancy is distributed equally across all dimensions and each dimension still preserves task-relevant information, the encoded feature vector may have a high dimension that leads to a high communication overhead.
To obtain a feature vector $\hat{Z}$ that aggregates the task-irrelevant information into certain expendable dimensions through end-to-end training, we adopt the log-uniform distribution as the variational prior, i.e., $q(\boldsymbol{\hat{z}})$, to induce sparsity \cite{kingma2015variational}. 
In particular, we choose the mean-ﬁeld variational approximation \cite{kingma2013autovae} to alleviate the computation complexity, i.e., given an $n$-dimensional $\boldsymbol{\hat{z}}$, $q(\boldsymbol{\hat{z}}) = \prod_{i}^{n} q(\hat{z}_{i})$.
Specifically, for each dimension $\hat{z}_{i}$, the variational prior distribution is chosen as:
\begin{equation*}
    q\left(\log \left|\hat{z}_{i}\right|\right)=\mathrm{constant}.
\end{equation*}
Since $p_{\boldsymbol{\phi}}(\boldsymbol{\hat{z}}|\boldsymbol{x}) = \prod_{i}^{n} p_{\boldsymbol{\phi}}(\hat{z}_{i}|\boldsymbol{x}) $, the KL-divergence term in (\ref{loss_in_3}) can be decomposed into a summation:
\begin{equation}
\label{VIB_log_uniform}
D_{K L}\left(p_{\boldsymbol{\phi}}(\boldsymbol{\hat{z}} | \boldsymbol{x}) \| q(\boldsymbol{x})\right)= \sum_{i=1}^{n} D_{K L}\left(p_{\boldsymbol{\phi}}\left(\hat{z}_{i} |\boldsymbol{x}\right) \| q\left(\hat{z}_{i}\right)\right).
\end{equation}
Nevertheless, as the KL-divergence term in (\ref{VIB_log_uniform}) does not have a closed-form expression, we utilize the approximation proposed in \cite{molchanov2017variational} as follows:
\begin{align}
    & -D_{K L}\left(p_{\boldsymbol{\phi}}\left(\hat{z}_{i} |\boldsymbol{x}\right) \| q\left(\hat{z}_{i}\right)\right) = \nonumber \\
    = & \frac{1}{2} \log \alpha_{i}-\mathbb{E}_{\epsilon \sim \mathcal{N}\left(1, \alpha_{i}\right)} \log |\epsilon|+\mathrm{C}  \nonumber\\
    \approx & k_{1} S \left(k_{2}+k_{3} \log \alpha_{i}\right)  -0.5 \log \left(1+\alpha_{i}^{-1}\right) +\mathrm{C}, \label{KL_approx}
\end{align}
where
\begin{equation*}
    \alpha_{i}=\frac{\sigma^{2}}{z_{i}^{2}} \quad k_{1}=0.63576 \quad k_{2}=1.87320 \quad k_{3}=1.48695,
\end{equation*}
and $\mathrm{C}$ is a constant. 
Besides, $z_{i}$ is the $i$-th dimension in $\boldsymbol{z}$, and $S(\cdot)$ denotes the sigmoid function.
It can be verified that the approximate KL-divergence approaches its minimum when $\alpha_{i}$ goes to infinite (i.e., $z_{i}$ goes to zero), and minimizing this term encourages the value of $z_{i}$ to be small.
Empirical results in Section \ref{experiment} show that the selected sparsity-inducing distribution sparsifies some dimensions in $\boldsymbol{z}$, i.e., $z_{i} \equiv 0$ for arbitrary input, which can be pruned to reduce the communication overhead.

\begin{algorithm}[t]
\small
\caption{Training Variational Feature Encoding (VFE)}
\begin{algorithmic}[1]
\label{statics_algorithm}
\Require $T$ (number of iterations), $n$ (number of output dimension of encoder), $L$ (number of channel noise samples per datapoint), batch size $M$, channel variance $\sigma^{2}$, and threshold $\gamma_{0}$.
\While{epoch $t=1$ to $T$}
\State Select a mini-batch of data $\left\{\left(\boldsymbol{x_{m}}, \boldsymbol{y_{m}}\right)\right\}_{m=1}^{M}$
\State Compute the encoded feature vector $\left\{\boldsymbol{z_{m}}\right\}_{m=1}^{M}$ based on (\ref{mu_forward_basic})
\State Compute the appropriate KL-divergence based on (\ref{KL_approx})
\While{$m=1$ to $M$}
\State Sample the noise $\left\{ \boldsymbol{\epsilon}_{\boldsymbol{m},\boldsymbol{l}}\right\}_{l=1}^{L} \sim \mathcal{N}(\boldsymbol{0},\sigma^{2}\boldsymbol{I})$
\EndWhile
\State Compute the loss $\mathcal{L}_{VIB}(\boldsymbol{\phi},\boldsymbol{\theta})$ based on (\ref{loss_in_3_estimate})
\State Update parameters $\boldsymbol{\phi},\boldsymbol{\theta}$ through backpropagation.
\While{$i=1$ to $n$}
        \If {$\gamma_{i} \leq \gamma_{0}$}
                \State Prune the $i$-th dimension in the encoded feature vector
        \EndIf
\EndWhile
\EndWhile
\end{algorithmic}
\end{algorithm}

\subsection{Variational Pruning on Dimension Importance}
While the selected variational prior helps to promote sparsity in the feature vector, we still need an effective method to determine which of the dimensions can be pruned.
Maintaining $z_{i} \equiv 0$ requires all the weights and the bias corresponding to $z_{i}$ in this layer to converge to zero.
However, checking each parameter is time-consuming in a large-scale DNN.
To develop an efficient solution, we introduce a \textit{dimension importance} vector $\bm{\gamma}$ to denote the importance of each output neuron.
Revisiting the fully-connected (FC) layer, each neuron has full connections to its input $\bm{a}$, and their activations can thus be computed with a matrix multiplication with $\bm{W}$ followed by an offset $\bm{b}$ as follows:
\begin{equation}
\begin{aligned}
    \boldsymbol{\mathcal{FC}({a})} = & \boldsymbol{Wa} + \boldsymbol{b} = \boldsymbol{\widetilde{W}\tilde{a}},
\end{aligned}
\label{fc_layer}
\end{equation}
where $\boldsymbol{\widetilde{W}} = \boldsymbol{[W,b]}$ is an augmented weight matrix, and $\boldsymbol{\tilde{a}}=[\boldsymbol{a}^{T},1]^{T}$ is an augmented input vector.
By denoting the $i$-th row in the augmented weight matrix $\boldsymbol{\widetilde{W}}$ as $\bm{\widetilde{W}_{i\cdot}}$ and the $i$-th dimension in $\boldsymbol{\gamma}$ as $\gamma_{i}$, we rewrite the augmented weight matrix as $\bm{\widetilde{W}_{i\cdot}} = \gamma_{i} \frac{\bm{\widetilde{W}_{i\cdot}}}{\|\bm{\widetilde{W}_{i\cdot}}\|_{2}} $, where $\bm{\gamma}$ corresponds to the scale factor for each row.
The proposed VFE method defines the mapping from the input $\bm{x}$ to the encoded feature $\bm{z}$ according to the
following formula:

\begin{equation}
\label{mu_forward_basic}
z_{i} = \text{Tanh} \left( \gamma_{i}\frac{\boldsymbol{\widetilde{W}_{i\cdot}}}{\|\boldsymbol{\widetilde{W}_{i\cdot}}\|_{2}}\boldsymbol{f(x)}\right),
\end{equation}
where $z_{i}$ is the $i$-th dimension of $\bm{z}$, and $\text{Tanh}(\cdot)$ is the activation function. Besides, function $\boldsymbol{f(\cdot)}$ is defined by the previous on-device layers, and its output $\boldsymbol{f(x)}$ is the input of the fully-connected layer (i.e., $\bm{a} = \boldsymbol{f(x)}$
in (\ref{fc_layer})).
As the weight vector $\boldsymbol{\widetilde{W}_{i\cdot}}$ is normalized by its $l_{2}$-norm, the magnitude of $z_{i}$ is highly dependent on the scale factor $\gamma_{i}$.
When $\gamma_{i}$ is close to zero, $z_{i}$ is also close to zero, and the corresponding $p_{\boldsymbol{\phi}}(\boldsymbol{\hat{z}} |\boldsymbol{x})$ degrades to the channel noise distribution without valid information.
Based on this idea, we eliminate the redundant channels when the parameter $\gamma_{i}$ is less than a threshold $\gamma_{0}$.
Since the Tanh activation function has an output range from -1 to 1, the peak transmitted power $P$ is constrained to 1.
Note that the formula in (\ref{mu_forward_basic}) can be easily extended to convolutional layers by replacing the matrix multiplication with convolution.
Such a variational pruning process is one of the main components of the proposed VFE method.
The training procedures for VFE are illustrated in Algorithm 1.

\section{Variable-length Variational Feature Encoding}
\label{variable_length}

The task-oriented communication scheme developed in Section \ref{basic_method} assumes static wireless channels.
In practice, wireless data transmission may experience changes due to various factors such as beam blockage and signal attenuation.
This necessitates instant link adaptation to improve the efficiency of feature encoding for low-latency inference.
In this section, we extend our findings in Section \ref{basic_method} and propose a new encoding scheme, namely variable-length variational feature encoding (VL-VFE), by designing a dynamic neural network, which admits flexible control of the encoded feature dimension.

\subsection{Background on Dynamic Neural Networks}

\begin{figure*}[t]
\centering
\subfloat[Random activation]{
\label{random_activation}
\centering
\includegraphics[width=0.85\linewidth]{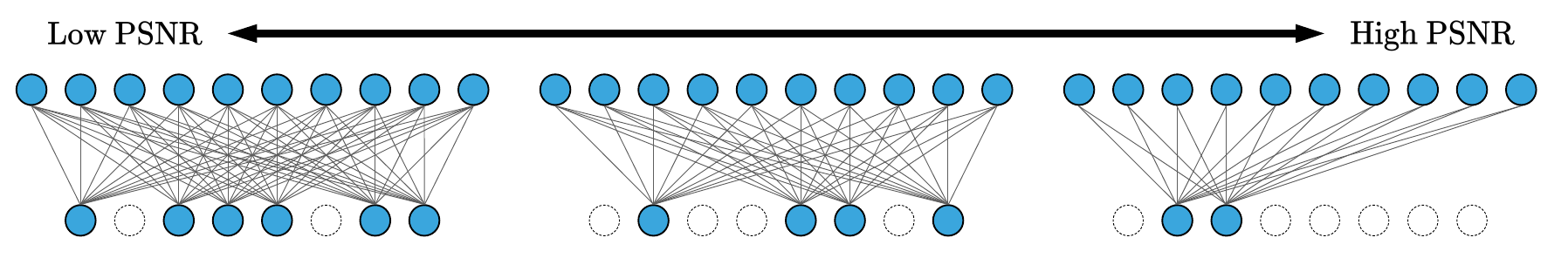}
}
\vfill
\subfloat[Consecutive activation]{
\centering
\label{consecutive_activation}
\includegraphics[width=0.85\linewidth]{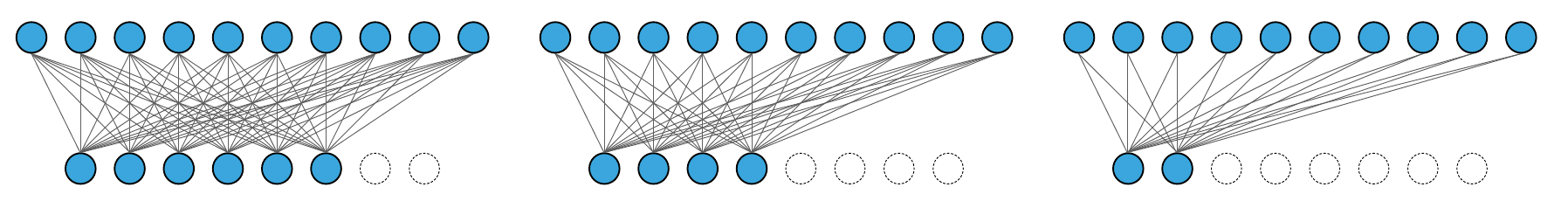}
}

\caption{Two types of selective activations: Random activation and consecutive activation. In different channel conditions (e.g., different PSNRs), the same DNN can be executed with different activated dimensions to balance the achievable inference performance and the incurred communication overhead. Random activation does not require the dimensions to be activated in order, while consecutive activation forces the activated dimensions to be consecutive starting from the first dimension.}
\label{selective_activation}
\end{figure*}

Dynamic neural networks are able to adapt their architectures to the given input and are effective in improving the efficiency of the network processing via selective execution.
For example, several prior works (e.g. \cite{wang2018skipnet,wu2018blockdrop,chen2019you_gaternet}) proposed to learn a binary gating module to adaptively skip layers or prune channels based on the input data.
Besides, there are also some variants of dynamic neural networks, including the slimmable neural networks and the ``\emph{Once-for-All}'' architecture. In particular, inventors of the slimmable neural networks \cite{yu2019slimmable} proposed to train a single model to support layers with arbitrary widths; while authors of \cite{cai2019once} proposed the ``Once-for-All'' architecture with a progressive shrinking algorithm that trains one network to support diverse sub-networks.
In this work, we employ the idea of selective activation, as shown in Fig. \ref{selective_activation}, to learn a set of neurons that can adjust the number of activated neurons according to the channel conditions.

\subsection{Selective Activation for Dynamic Channel Conditions}
We propose the variable-length variational feature encoding (VL-VFE), which is empowered with the capability of adjusting its output length under different channel conditions.
Such kinds of channel-adaptive feature encoding schemes favor the following two properties:
\begin{itemize}
\item The activated dimensions of the feature $\boldsymbol{z}$ can be adjusted in the DNN forward propagation according to the channel conditions. More dimensions should be activated during the bad channel conditions and vice versa.
\item The activated dimensions start consecutively from the first dimension (shown in Fig. \ref{consecutive_activation}), which avoids transmitting the indexes of the activated dimensions using extra communication resources.
\end{itemize}

In practical communication systems, the mobile device could be aware of the channel condition via a feedback channel.
Therefore, the channel condition can be incorporated in the feature encoding process.
Because the amplitude of the encoded feature vector is constrained to 1 by Tanh function, the noise variance $\sigma^{2}$ suffices to represent the PSNR and is adopted as an extra input of the feature encoder.
In the training process, the noise variance $\sigma^{2}$ is regarded as a random variable distributed within a range to model the dynamic channel conditions.
For simplicity, we sample the channel variance $\sigma^{2}$ from the uniform distribution $p(\sigma^{2})$.
As the noise variance $p(\sigma^{2})$ is independent to the dataset, we have $p(\boldsymbol{x},\boldsymbol{y},\sigma^{2}) = p(\boldsymbol{x},\boldsymbol{y})p(\sigma^{2})$.
The loss function in (\ref{loss_in_3}) is thus revised as follows:
\begin{equation}
\label{loss_in_4}
\begin{aligned}
    \widetilde{\mathcal{L}}_{VIB}(\boldsymbol{\phi},\boldsymbol{\theta}) = & \ \mathbf{E}_{p(\boldsymbol{x},\boldsymbol{y},\sigma^{2})} \left\{ \mathbf{E}_{p_{\boldsymbol{\phi}}(\boldsymbol{\hat{z}} | \boldsymbol{x},\sigma^{2})}\left[-\log q_{\boldsymbol{\theta}}(\boldsymbol{y} | \boldsymbol{\hat{z}})\right] \right. \\
    & +  \beta D_{K L}\left(p_{\boldsymbol{\phi}}(\boldsymbol{\hat{z}} | \boldsymbol{x},\sigma^{2}) \| q(\boldsymbol{\hat{z}})\right) \Big\}.
\end{aligned}
\end{equation}
Similarly, we adopt Monte Carlo sampling as in (\ref{loss_in_3_estimate}) to estimate $\widetilde{\mathcal{L}}_{VIB}$. 
The formula is as follows:
\begin{equation}
\label{loss_in_4_estimate}
\begin{aligned}
\widetilde{\mathcal{L}}_{VIB}(\boldsymbol{\phi},\boldsymbol{\theta}) \simeq & \frac{1}{M} \sum_{m=1}^{M}\left\{ \frac{1}{L} \sum_{l=1}^{L} \left[-\log q_{\boldsymbol{\theta}}\left(\boldsymbol{y_{m}} |\boldsymbol{\hat{z}}_{\boldsymbol{m},\boldsymbol{l}}\right)\right] \right. \\
& + \beta D_{K L}\left(p_{\boldsymbol{\phi}}(\hat{\boldsymbol{z}} | \boldsymbol{x}_{\boldsymbol{m}},\sigma_{m}^{2}) \| q(\hat{\boldsymbol{z}})\right)\Bigg\},
\end{aligned}
\end{equation}
where $\boldsymbol{\hat{z}}_{\boldsymbol{m},\boldsymbol{l}} = \boldsymbol{z_{m}} + \boldsymbol{\epsilon}_{\boldsymbol{m},\boldsymbol{l}}$, $\sigma^{2}_{m} \sim p(\sigma^{2})$, and $\boldsymbol{\epsilon}_{\boldsymbol{m},\boldsymbol{l}} \sim \mathcal{N}(\boldsymbol{0},\sigma^{2}_{m}\boldsymbol{I})$, and for a given $\boldsymbol{z_{m}}$, the channel noise is sampled $L$ times.
Then, as the encoding scheme should be channel-adaptive, we have $p_{\boldsymbol{\phi}}\left(\hat{\boldsymbol{z}} | \boldsymbol{x},\sigma^{2}\right) = \mathcal{N}(\bm{\hat{z}}|\boldsymbol{z}(\boldsymbol{x};\boldsymbol{\phi},\sigma^{2}),\sigma^{2}\boldsymbol{I})$, where the function $\boldsymbol{z}(\boldsymbol{x};\boldsymbol{\phi},\sigma^{2})$ determined by the on-device network incorporates $\sigma^{2}$ as an input variable.
Hence, the function in (\ref{mu_forward_basic}) is modified as follows:
\begin{equation}
\label{on_device_forward_dynamic}
    z_{i}=\text{Tanh} \left(\gamma_{i}(\sigma^{2})\frac{{\boldsymbol{\widetilde{W}_{i\cdot}}}}{\|{\boldsymbol{\widetilde{W}_{i\cdot}}}\|_{2}}  \boldsymbol{f(}\boldsymbol{x}\boldsymbol{)}\right),
\end{equation}
where the dimension importance $\gamma_{i}(\sigma^{2})$ (i.e., the $i$-th element in $\boldsymbol{\gamma(}\sigma^{2}\boldsymbol{)}$) is a function of the channel condition (i.e., channel noise variance $\sigma^{2}$). 
Rather than directly training a gating network to control the activated dimensions like other dynamic neural networks (e.g., \cite{wang2018skipnet,wu2018blockdrop,chen2019you_gaternet}), $\boldsymbol{\gamma(}\sigma^{2}\boldsymbol{)}$ can adaptively prune the redundant dimensions in the encoded feature vector for different $\sigma^{2}$ due to the intrinsic sparsity discussed in Section \ref{basic_method}.
As a result, in the device-edge co-inference system, the activated dimensions of the encoded feature vector can be easily decided by setting a threshold for $\boldsymbol{\gamma(}\sigma^{2}\boldsymbol{)}$.
Besides, as VL-VFE needs to meet the consecutive activation property, we define the function $\boldsymbol{\gamma(}\sigma^{2}\boldsymbol{)}$ to induce a particular group sparsity pattern, and for the $i$-th element $\gamma_{i}(\sigma^{2})$, the expression is constructed as follows:
\begin{equation}
\label{gamma_sigma_compute}
    \gamma_{i}(\sigma^{2}) =\sum_{j=i}^{n} g_{j}(\sigma^{2}),
\end{equation}
where $g_{j}(\cdot)$ denotes the $j$-th output dimension of the function $\boldsymbol{g(}\cdot \boldsymbol{)}$, which is parameterized by a lightweight multi-layer perceptron (MLP).
By constraining the range of parameters in the MLP, each function $g_{j}(\sigma^{2})$ can be a non-negative increasing function, which naturally leads to $\gamma_{i}(\sigma^{2}) \geq \gamma_{j}(\sigma^{2}), \forall j>i$ and $\gamma_{i}(\sigma^{2}) \geq \gamma_{i}(\bar{\sigma}^{2}), \forall \sigma^{2} \geq \bar{\sigma}^{2}$.
Therefore, given a threshold $\gamma_{0}$, the VL-VFE method summarized in Algorithm 2 can activate the dimensions consecutively, and more dimensions can be activated during the adverse channel conditions.
Details of the MLP structure and parameter constraints are deferred to Appendix \ref{appendix b}.

\begin{algorithm}[t]
\small
	\caption{Training Variable-Length Variational Feature Enoding (VL-VFE)}  
	\begin{algorithmic}[1]
	\label{dynamic_algorithm}
\Require $T$ (number of iterations), $n$ (number of output dimension of encoder), $L$ (number of channel noise samples per datapoint), batch size $M$, noise variance distribution $p(\sigma^{2})$, and threshod $\gamma_{0}$.
\While{epoch $t=1$ to $T$}
\State Get a mini-batch of data $\left\{\left(\boldsymbol{x_{m}}, \boldsymbol{y_{m}}\right)\right\}_{m=1}^{M}$
\State Sample the channel variance $\left\{\sigma_{m}^{2}\right\}_{m=1}^{M} \sim p(\sigma^{2})$
\State Compute the encoded feature vector $\left\{\boldsymbol{z_{m}}\right\}_{m=1}^{M}$ based on (\ref{on_device_forward_dynamic})
\While{$m=1$ to $M$}
\State Sample the channel noise $\left\{\boldsymbol{\epsilon}_{\boldsymbol{m},\boldsymbol{l}}\right\}_{l=1}^{L} \sim \mathcal{N}(0,\sigma_{m}^{2}\boldsymbol{I})$
\While{$i=1$ to $n$}
        \If {$\gamma_{i}(\sigma^{2}_{m}) \leq \gamma_{0}$}
                \State Deactivate the $i$-th dimension of $\boldsymbol{z}_{m}$ in this epoch
        \EndIf
\EndWhile
\EndWhile
\State Compute the appropriate KL-divergence based on (\ref{KL_approx})
\State Compute loss $\widetilde{\mathcal{L}}_{VIB}(\boldsymbol{\phi},\boldsymbol{\theta}) $ based on (\ref{loss_in_4_estimate})
\State Update parameters $\boldsymbol{\phi},\boldsymbol{\theta}$ through backpropagation
\EndWhile
\end{algorithmic}
\end{algorithm}

\subsection{Training Procedure for the Dynamic Neural Network}

To train a dynamic neural network with the selective activation under different channel conditions, we naturally average losses sampled from different cases.
In each training iteration, for simplicity, we sample $\sigma^{2}$ from the possible PSNR range.
Different from the training procedure in Algorithm 1, VL-VFE deactivates each dimension with $\gamma_{i}\left(\sigma^{2}\right) \leq \gamma_{0}$, rather than permanently pruning it, as the function $\boldsymbol{\gamma(}\sigma^{2}\boldsymbol{)}$ is not stable until convergence.
More details about the algorithm are summarized in Algorithm 2.

\section{Performance Evaluation}
\label{experiment}

In this section, we evaluate the performance of the proposed task-oriented communication schemes on image classification tasks and investigate the rate-distortion tradeoff for both static and dynamic channel conditions.
An ablation study is also conducted to illustrate the importance of an appropriate choice of the variational prior distribution discussed in Section \ref{basic_method}, i.e., a sparsity-inducing prior distribution can force some dimensions of the encoded feature vector to zero without over-shrinking other dimensions.

\subsection{Experimental Setup}
\subsubsection{Datasets} In this section, we select two benchmark datasets for image classification, including MNIST \cite{lecun1998gradientMNIST} and CIFAR-10 \cite{cifar100}.
The MNIST dataset of handwritten digits from ``0'' to ``9'' has a training set of 60,000 sample images and a test set of 10,000 sample images. The CIFAR-10 dataset consists of 60,000 color images in 10 classes with 5,000 training images per class and 10,000 test images.
{In Appendix \ref{appendix D}, we further test the performance of the proposed methods on the Tiny Imagenet dataset \cite{tiny_imagenet}.}
\subsubsection{Baselines}
We compare the proposed methods against two learning-based communication schemes for device-edge co-inference, including \textbf{DeepJSCC} \cite{jssc_deniz,jankowski2020deep} and \textbf{learning-based} \textbf{Quantization} \cite{hubara2017quantized}.
\begin{itemize}
\item \textbf{DeepJSCC}: {DeepJSCC is a learning-based JSCC method, which maps the input data directly to the channel symbols via a JSCC encoder. We set the loss function of DeepJSCC to cross-entropy, and its communication cost is proportional to the output dimension of the feature encoder.}
\item \textbf{Learning-based Quantization}: This scheme quantizes the floating-point values in the encoded feature vector into low-precision data representations (e.g., the 2-bit fixed-point format). Such a quantization method imitates the lossy source coding and therefore it requires an extra step of channel coding before transmission for error correction. Note that designing a universally optimal channel coding scheme for different channel conditions in the finite block-length regime is highly nontrivial \cite{finite_channel_coding}. For fair comparisons, we assume an adaptive channel coding scheme that achieves the following communication rate:
\begin{equation}
\label{capability_upper}
\begin{aligned}
    & C(P,\sigma^{2}) = \\ 
    = & \min \left\{\log_{2} \left(1+\sqrt{ \frac{2P}{\pi e \sigma^{2}}}\right), \frac{1}{2} \log_{2} \left(1+\frac{P}{\sigma^{2}}\right)\right\} \text{(b.p.c.u)},
\end{aligned}
\end{equation}
where $\frac{P}{\sigma^{2}}$ is the PSNR.
This formula was shown to be a tight upper bound on the capacity of the amplitude-limited scalar Gaussian channel in \cite{mckellips2004simple}.
\end{itemize}

\subsubsection{Metrics}
We mainly concern the rate-distortion tradeoff in task-oriented communication.
For the classification tasks, we use the classification accuracy to denote the inference performance (corresponding to ``distortion''), and adopt the communication latency as an indicator of ``rate''.
In the following experiments, we set the bandwidth $W$ as 12.5kHz with a symbol rate of 9,600 Baud, corresponding to the limited bandwidth at the wireless edge.

\subsubsection{Neural Network Architecture}
Carefully designing the on-device network is important due to the limited onboard computation and memory resources.
Besides, as the DNN structure affects the inference performance and communication overhead, all methods adopt the same architecture for fair comparisons as follows\footnote{The code is available at {github.com/shaojiawei07/VL-VFE}.}.
\begin{table}[]
\selectfont
\begin{center}
\caption{The DNN structure for MNIST classification task.}
\resizebox{0.475\textwidth}{!}{
\begin{tabular}{c|c|c}
\hline
                                      & \textbf{Layer}                           & \begin{tabular}[c]{@{}c@{}}\textbf{Output} \\ \textbf{Dimensions}\end{tabular} \\ \hline
\begin{tabular}[c]{@{}c@{}}\textbf{On-device} \\ \textbf{Network}\end{tabular}                   & Fully-connected Layer + Tanh    & $n$                 \\ \hline
\multirow{3}{*}{\begin{tabular}[c]{@{}c@{}}\textbf{Server-based} \\ \textbf{Network}\end{tabular}} & Fully-connected Layer + ReLU    & 1024              \\ \cline{2-3} 
                                      & Fully-connected Layer + ReLU    & 256               \\ \cline{2-3} 
                                      & Fully-connected Layer + Softmax & 10                \\ \hline
\end{tabular}
}
\label{network_structure_mnist}
\end{center}
\end{table}

\begin{table}[]
\selectfont
\begin{center}
\caption{The DNN structure for CIFAR-10 classification task.}
\resizebox{0.475\textwidth}{!}{
\begin{tabular}{c|c|c}
\hline
                                      & \textbf{Layer}                                  & \begin{tabular}[c]{@{}c@{}}\textbf{Output} \\ \textbf{Dimensions}\end{tabular} \\ \hline
\multirow{4}{*}{\begin{tabular}[c]{@{}c@{}}\textbf{On-device} \\ \textbf{Network}\end{tabular}}    & {[}Convolutional Layer + ReLU{]} $\times$ 2   & 128 $\times$ 32 $\times$ 32     \\ \cline{2-3} 
                                      & ResNet Building Block                  & 128 $\times$ 16 $\times$ 16     \\ \cline{2-3} 
                                      & {[}Convolutional Layer + ReLU{]} $\times$ 2   & 4 $\times$ 4 $\times$ 4         \\ \cline{2-3} 
                                      & Reshape + Fully-connected Layer + Tanh & n                 \\ \hline
\multirow{5}{*}{\begin{tabular}[c]{@{}c@{}}\textbf{Server-based} \\ \textbf{Network}\end{tabular}} & Fully-connected Layer + ReLU + Reshape & 64                \\ \cline{2-3} 
                                      & {[}Convolutional Layer + ReLU{]} $\times$ 2   & 512 $\times$ 4 $\times$ 4       \\ \cline{2-3} 
                                      & ResNet Building Block                  & 512 $\times$ 4$ \times$ 4       \\ \cline{2-3} 
                                      & Pooling Layer                          & 512               \\ \cline{2-3} 
                                      & Fully-connected Layer + Softmax        & 10                \\ \hline
\end{tabular}
}
\label{network_structure_cifar-10}
\end{center}
\end{table}

\begin{figure*}[t]
\centering
\subfloat[PSNR = 10 dB]{
\centering
\includegraphics[width=0.4\linewidth]{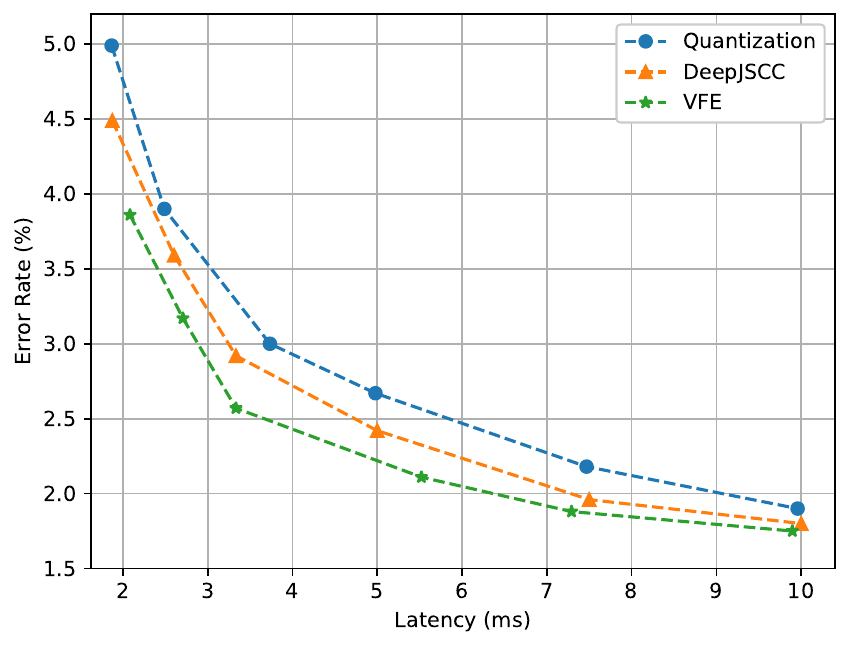}
}
\subfloat[PSNR = 20 dB]{
\centering
\includegraphics[width=0.4\linewidth]{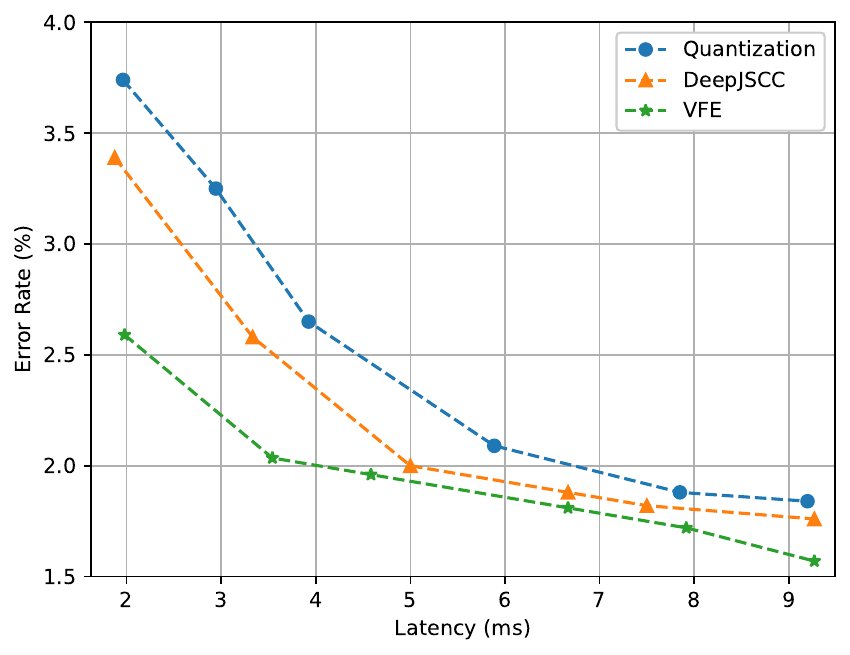}
}

\caption{The rate-distortion curves in the MNIST classification task with (a) PSNR = 10 dB and (b) PSNR = 20 dB.}
\label{exp_RD_MNIST}

\end{figure*}

\begin{figure*}[t]
\centering
\subfloat[PSNR = 10 dB]{
\centering
\includegraphics[width=0.4\linewidth]{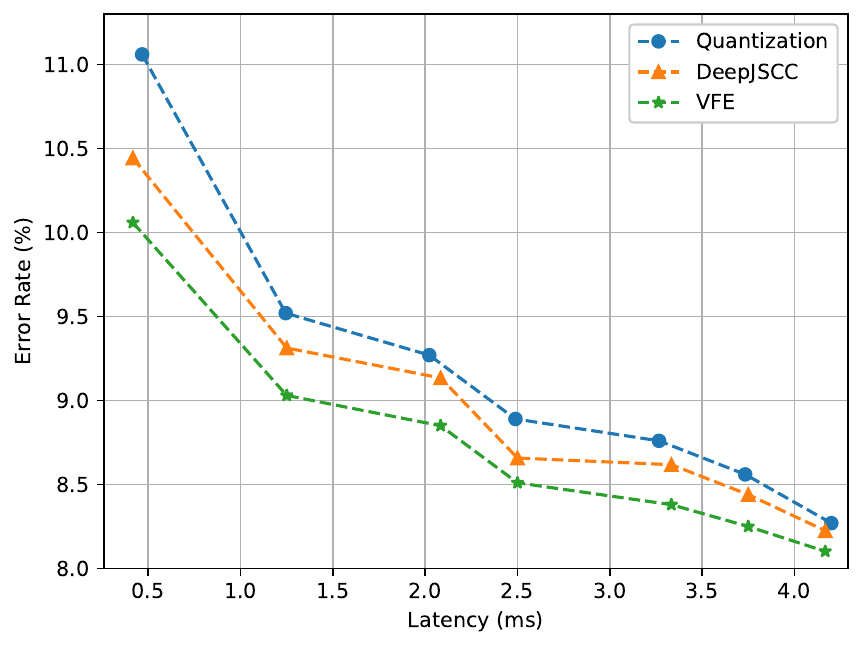}
}
\subfloat[PSNR = 20 dB]{
\centering
\includegraphics[width=0.393\linewidth]{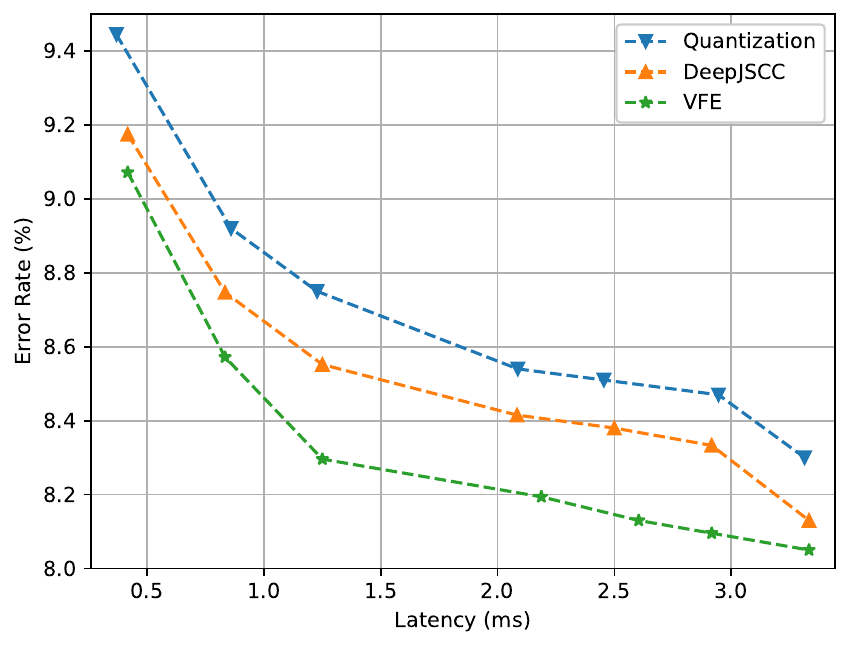}
}

\caption{The rate-distortion curves in the CIFAR-10 classification task with (a) PSNR = 10 dB and (b) PSNR = 20 dB.}
\label{exp_RD_CIFAR}

\end{figure*}

\begin{itemize}
\item For the MNIST classification experiment, we assume a microcontroller unit (e.g., ARM STM32F4 series) as the mobile device, and its memory (RAM) is less than 0.5 MB.
Therefore, we use only one fully-connected layer as the on-device network to meet the memory constraint.
At the edge server, we select an MLP as the server-based network.
The corresponding network structure is shown in Table \ref{network_structure_mnist}.
Note that a 4-layer MLP achieves an error rate of 1.38\% as reported in \cite{alemi2016deepVIB}.
\item For the CIFAR-10 classification task, we assume a single-board computer (e.g., Raspberry Pi series) as the mobile device and adopt ResNet \cite{ResNet} as the backbone for the CIFAR-10 processing, which can achieve the classification accuracy of around 92\%.
As the single-board computer has much more memory compared to a microcontroller, we deploy convolutional layers on the mobile device to extract a compact representation.
Besides, to reduce the communication overhead, we add a fully-connected layer at the end of the on-device network to map the intermediate tensor to an $n$-dimensional encoded feature.
Correspondingly, there is a fully-connected layer in the server-based network that maps the received feature vector back to a tensor, and several server-based layers are adopted for further processing.
The network structure is shown in Table \ref{network_structure_cifar-10}.
\end{itemize}

Since the proposed methods can prune the redundant dimensions in the encoded feature vector, our methods initialize $n$ to 64 or 128 in the following experiments.
Moreover, the function $\boldsymbol{g(}\cdot\boldsymbol{)}$ in (\ref{gamma_sigma_compute}) for variable-length encoding is a 3-layer MLP with 16 hidden units each, which brings negligible computation compared with other computation-intensive modules\footnote{Note that there is a tradeoff between the on-device computation latency and the communication overhead caused by the complexity of the on-device network \cite{shao2020communication}.
In this paper, as we assume an extreme bandwidth-limited situation, we mainly consider the communication overhead in the experiments.}.

\begin{figure*}[t]
\centering
\subfloat[DeepJSCC: Accuracy = 96.77\%, dimension $n$ = 24.]{
\centering
\includegraphics[width=0.4\linewidth]{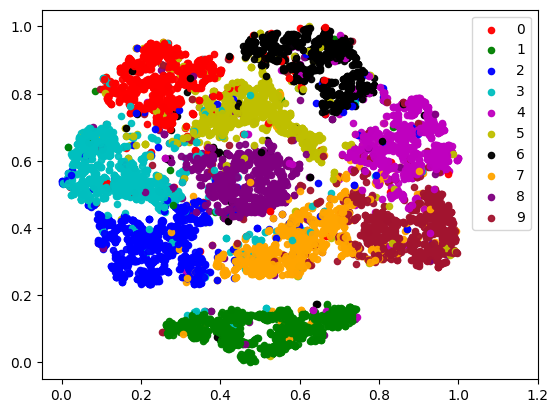}
}
\subfloat[Proposed VFE: Accuracy = 97.39\%, dimension $n$ = 24.]{
\centering
\includegraphics[width=0.4\linewidth]{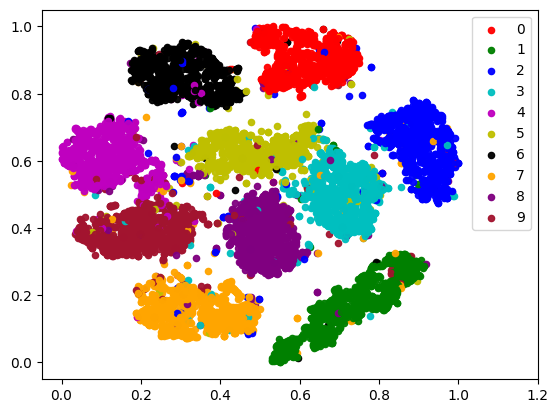}
}

\caption{2-dimensional t-SNE embedding of the received feature in the MNIST classification task with PSNR = 20 dB.}
\label{conste_mnist}

\end{figure*}

\subsection{Results for Static Channel Conditions}

In this set of experiments, we assume the wireless channel model has the same value of PSNR in both the training and test phases.
Then, we record the inference accuracy achieved with different communication latency to obtain the rate-distortion tradeoff curves. 
In the proposed VFE method, varying the weighting parameter $\beta$ can adjust the encoded feature length, where $\beta \in [10^{-4},10^{-2}]$ in the MNIST classification, and $\beta \in [5\times10^{-5},10^{-2}]$ in the CIFAR-10 classification.
The communication latency of DeepJSCC is determined by the encoded feature dimension $n$, while for the learning-based Quantization method, the communication latency is determined by the dimension $n$ and the number of quantization levels.
Adjusting these parameters affects both the communication latency and accuracy.
The rate-distortion tradeoff curves are shown in Fig. \ref{exp_RD_MNIST} and Fig. \ref{exp_RD_CIFAR} for the MNIST and CIFAR-10 classification tasks, respectively. It shows that our proposed method outperforms the baselines by achieving a better rate-distortion tradeoff, i.e., with a given latency requirement, a higher classification accuracy is maintained, and vice versa. This is because the proposed VFE method is able to identify and eliminate the redundant dimensions of the encoded feature vector for the task-oriented communication. 
Besides, we also depict the noisy feature vector $\boldsymbol{\hat{z}}$ in the MNIST classification tasks in Fig. \ref{conste_mnist} using a two-dimensional t-distributed stochastic neighbor embedding (t-SNE) \cite{maaten2008visualizing}.
Since the IB principle can preserve less nuisance from the input and make $\boldsymbol{\hat{z}}$ less affected by the channel noise, our VFE method can better distinguish the data from different classes compared with DeepJSCC.

\begin{table}[]
\selectfont
\caption{The classification accuracy under different PSNR with communication latency $t \leq 3.25 \ $ ms. The baseline classifiers achieve an accuracy of 98.64\% for MNIST classification and an accuracy of around  92\% for CIFAR-10 classification.}
\begin{center}
\resizebox{0.465\textwidth}{!}{
\begin{tabular}{c|cccc}
\hline \textbf{MNIST} & 10 dB & 15 dB & 20 dB & 25 dB \\
\hline DeepJSCC & 97.04 & 97.13 & 97.45 & 97.56 \\
 Quantization & 95.32 & 95.96 & 96.81 & 97.12 \\
\textbf{Proposed} & \textbf{97.29} & \textbf{97.79} & \textbf{98.01} & \textbf{98.17} \\
\hline \textbf{CIFAR-10} & 10 dB & 15 dB & 20 dB & 25 dB \\
\hline DeepJSCC & 91.58 & 91.60 & 91.67 & 91.72 \\
 Quantization & 90.68 & 91.07 & 91.53 & 91.65 \\
\textbf{Proposed} & \textbf{91.62} & \textbf{91.72} & \textbf{91.90} & \textbf{92.04} \\
\hline
\end{tabular}
}
\label{table-awgn}
\end{center}
\end{table}

Next, we test the robustness of the proposed method by further evaluating its inference performance over different channel conditions.
Particularly, we set a transmission latency tolerance of 3.25 ms and record the best inference accuracy achieved by different schemes\footnote{Theoretically, based on the channel capacity bound in (\ref{capability_upper}), transmitting a MNIST image takes around 8 ms when PSNR = 25 dB and 20 ms when PSNR = 10 dB.
Similarly, transmitting a CIFAR-10 image takes around 70 ms when PSNR = 25 dB and 180 ms when PSNR = 10 dB.}.
Since the channel achievable rate decreases with the PSNR, it requires the learning-based Quantization method to reduce the encoded data size to meet the latency constraint.
The latency constraint can also be translated to an encoded feature vector with less than 32 dimensions for both the VFE method and DeepJSCC.
Table \ref{table-awgn} shows the classification accuracy under various values of PSNR for the MNIST and CIFAR-10 tasks.
It is observed that, our method consistently outperforms the two baselines, implying that the IB framework can effectively identify the task-relevant information in the encoding scheme, and our VFE method is capable of achieving resilient transmission for task-oriented communication. 

\subsection{Results for Dynamic Channel Conditions}

In this subsection, we evaluate the performance of the proposed VL-VFE method in dynamic channel conditions. We assume the PSNR is changing from 10 to 25 dB. As the peak transmit power is constrained below 1 by the Tanh activation function, it equivalently means that the channel noise variance $\sigma^{2}$ varies in $[3\times10^{-3},0.1]$.
We compare the inference performance between the proposed method and DeepJSCC when testing in a wide range of PSNR.
The DeepJSCC is trained with PSNR = 20 dB, and its feature dimension is set to $n = 36$ in the MNIST classification task and $n=16$ in the CIFAR-10 classification task.

\begin{figure*}[t]
\centering
\subfloat[The MNIST classification task]{
\centering
\includegraphics[width=0.413\linewidth]{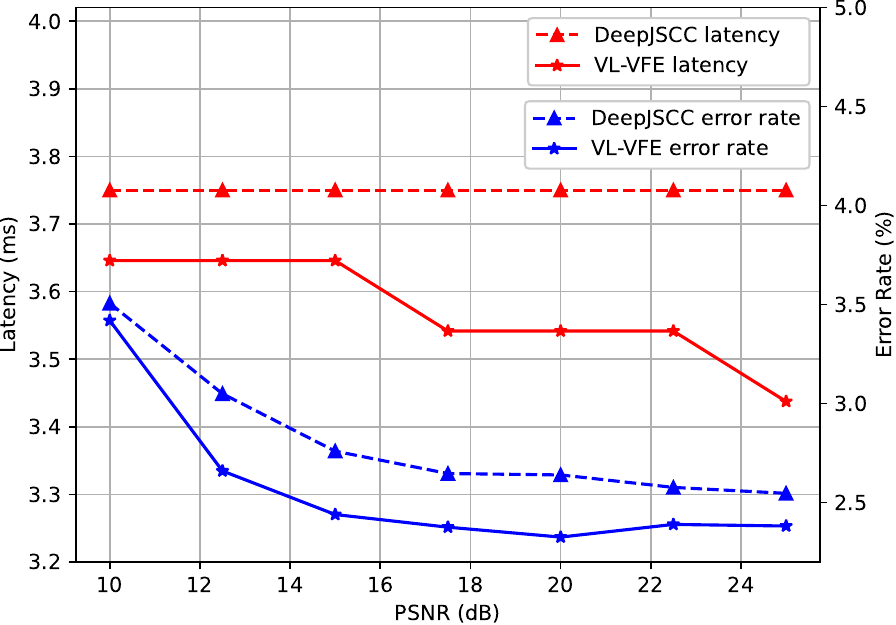}
}
\subfloat[The CIFAR-10 classification task]{
\centering
\includegraphics[width=0.422\linewidth]{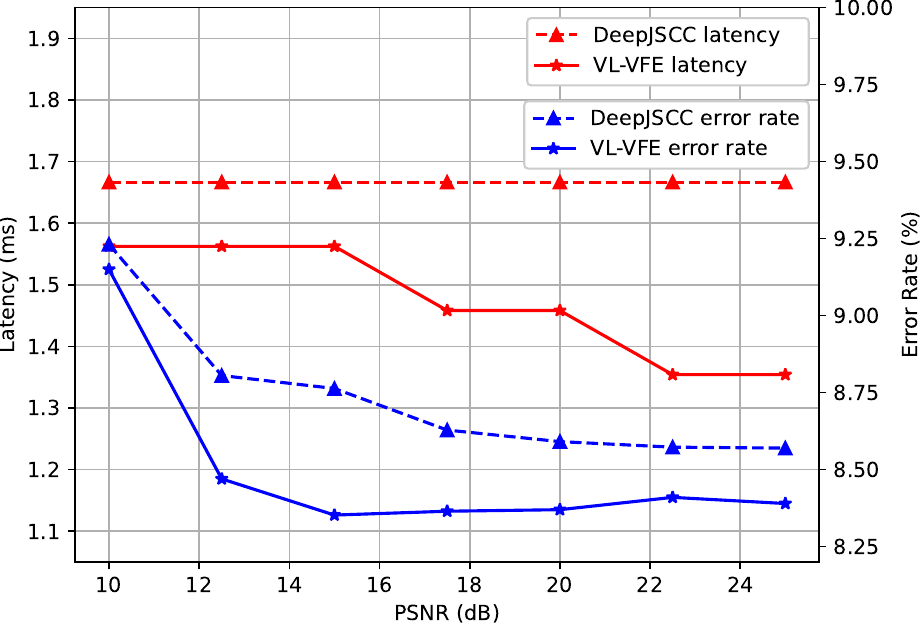}
}

\caption{Communication latency and error rate as a function of the channel PSNR in dynamic channel conditions.}
\label{exp_dynamic}
\end{figure*}

Fig. \ref{exp_dynamic} shows the latency and inference accuracy for the two classification tasks, which illustrates that the proposed VL-VFE method achieves a higher accuracy and lower latency compared with DeepJSCC.
The proposed VL-VFE method can adaptively adjust the encoded feature dimension according to the instantaneous channel noise level, and thus it can reduce the communication latency in the high PSNR regime.
Specifically, when the channel conditions are unfavorable, VL-VFE tends to activate more dimensions for transmission to make the received feature vector robust to maintain the inference performance, which is analogous to adding redundancy for error correction in conventional channel coding techniques.
On the contrary, when the channel conditions are good enough, VL-VFE tends to activate less dimensions to reduce the communication overhead.

Note that in existing communication systems, channel estimation plays a very important role in the performance of the whole system.
To evaluate the influence of the non-ideal estimation of the channel noise variance $\sigma^{2}$, we conduct the experiments to test the robustness of the proposed VL-VFE method given inaccurate noise variance $\hat{\sigma}^{2}$. More details of the experimental settings and results are deferred to Appendix \ref{appendix C}.

\begin{figure}[t] 
\centering
\subfloat[The $\boldsymbol{\gamma}$ value with a Gaussian distribution as the variational prior. Task accuracy = 95.91 \% with 21 activated dimensions.]{
\centering
\includegraphics[width=0.45\linewidth]{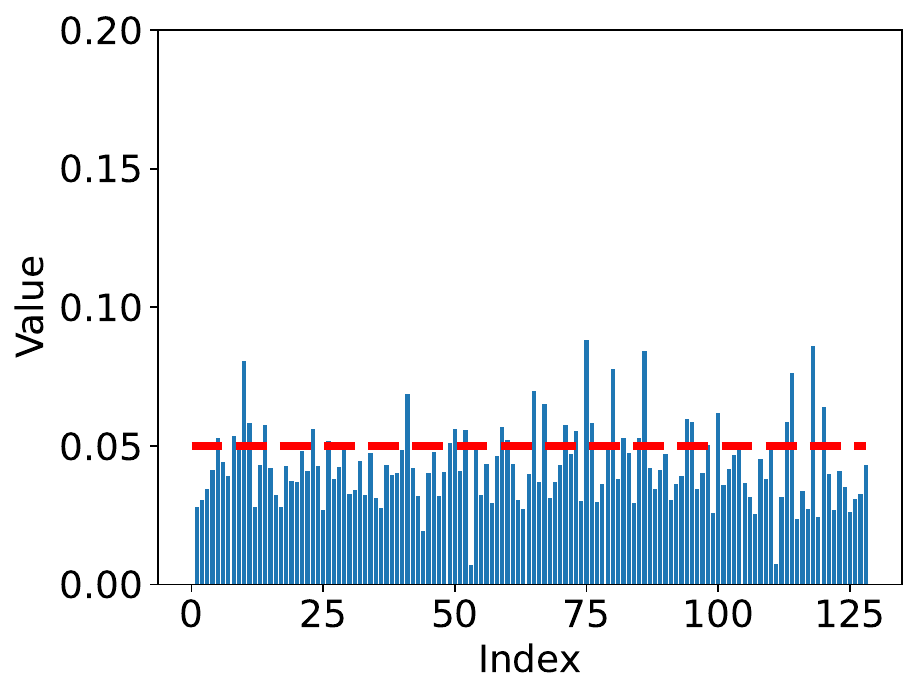}
}
\hspace{.1in}
\subfloat[The $\boldsymbol{\gamma}$ value with a log-uniform distribution as the variational prior. Task accuracy = 97.99 \% with 32 activated dimensions.]{
\centering
\includegraphics[width=0.45\linewidth]{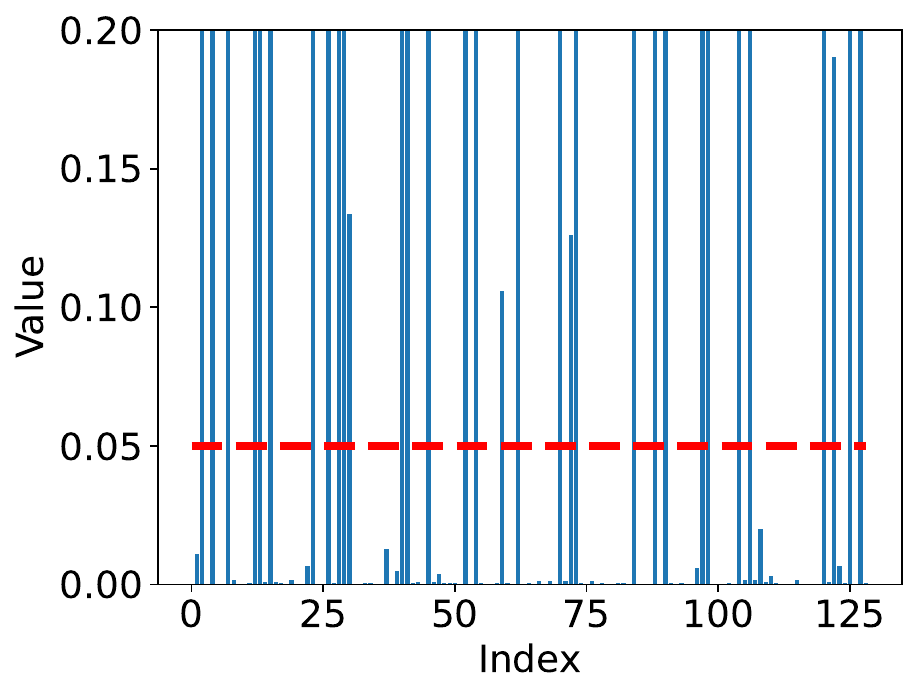}
}

\caption{The $\boldsymbol{\gamma}$ value in the MNIST classification task with (a) a Gaussian distribution as the variational prior and (b) a log-uniform distribution as the variational prior. The red dashed line denotes the pruning threshold $\gamma_{0}=0.05$.}
\label{gamma_mnist}

\end{figure}

\begin{figure}[ht]
\centering
\subfloat[The $\boldsymbol{\gamma}$ value with a Gaussian distribution as the variational prior. Task accuracy = 91.18 \% with 21 activated dimensions.]{
\centering
\includegraphics[width=0.45\linewidth]{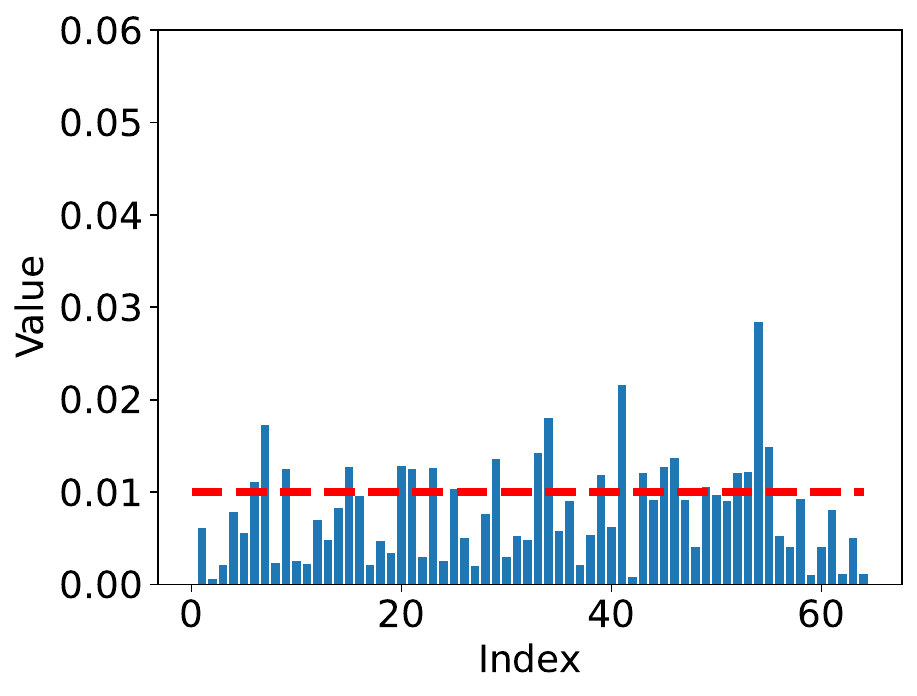}
}
\hspace{.1in}
\subfloat[The $\boldsymbol{\gamma}$ value with a log-uniform distribution as the variational prior. Task accuracy = 91.83 \% with 21 activated dimensions.]{
\centering
\includegraphics[width=0.45\linewidth]{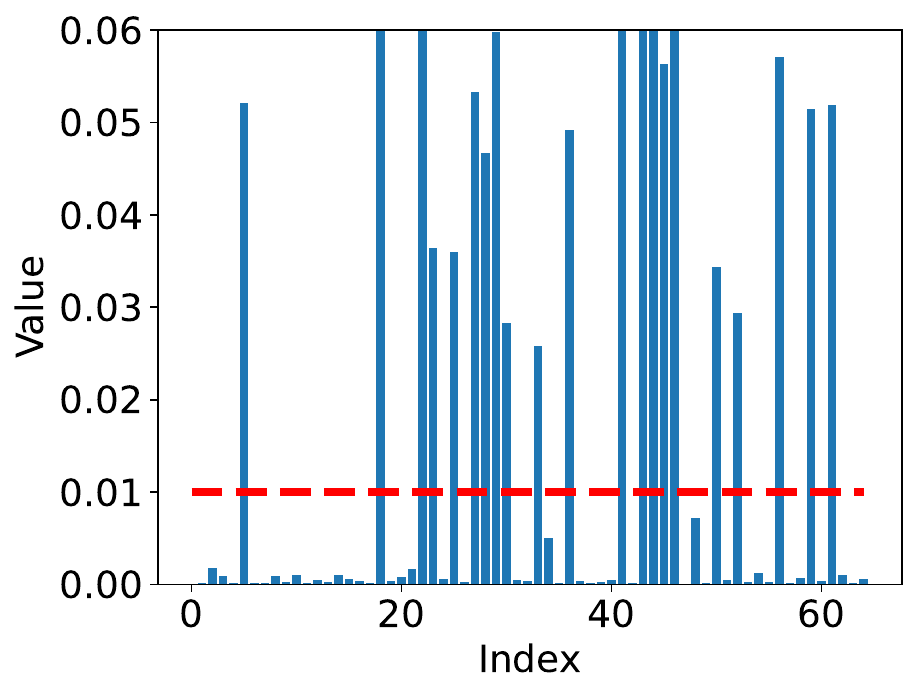}
}

\caption{The $\boldsymbol{\gamma}$ value in the CIFAR-10 classification task with (a) a Gaussian distribution as the variational prior and (b) a log-uniform distribution as the variational prior. The red dashed line denotes the pruning threshold $\gamma_{0}=0.01$.}
\label{gamma_cifar}

\end{figure}

\subsection{Ablation Study}

To verify the effectiveness of the log-uniform distribution as the variational prior $q(\boldsymbol{\hat{z}})$ for sparsity induction, we further conduct an ablation study that selects a Gaussian distribution with a diagonal covariance matrix for comparison.
Note that the Gaussian distribution is widely used in the previous variational approximation studies (e.g., \cite{kingma2013autovae,achille2018information_dropout}) as it generally has a closed-form solution.
Since the Gaussian distribution is not a parameter-free distribution, the mean value and covariance matrix are optimized in the training process to minimize the KL-divergence $D_{KL}(p(\boldsymbol{\hat{z}}|\boldsymbol{x})\|q(\boldsymbol{\hat{z}}))$.
The experiments are conducted for MNIST and CIFAR-10 classification assuming PSNR = 20 dB. 
The values of $\boldsymbol{\gamma}$ with different variational prior distributions are shown in Fig. \ref{gamma_mnist} and \ref{gamma_cifar}. 
The dashed line corresponds to the value of threshold $\gamma_{0}$ used to prune the dimensions.
From these two figures, it can be seen that, although using the Gaussian distribution can also confine some dimensions of $\boldsymbol{\gamma}$ to close-to-zero values, it is prone to shrinking the remaining informative dimensions that eventually results in inference accuracy degradation.

\begin{table*}
\centering
\begin{minipage}{0.99\textwidth}
\begin{equation}
\label{appendix:vib}
\begin{aligned}
 \mathcal{L}_{IB}(\boldsymbol{\phi}) = & -I(Y,\hat{Z})+\beta I(\hat{Z},X) \\
 = & - \int p(\boldsymbol{y}| \boldsymbol{\hat{z}})p(\boldsymbol{\hat{z}}) \log \frac{p(\boldsymbol{y} | \boldsymbol{\hat{z})}}{p(\boldsymbol{y})}  d \boldsymbol{y} d \boldsymbol{\hat{z}} + \beta \int p_{\boldsymbol{\phi}}(\boldsymbol{\hat{z}}|\boldsymbol{x})p(\boldsymbol{x}) \log \frac{p_{\boldsymbol{\phi}}(\boldsymbol{\hat{z}} | \boldsymbol{x)}}{p(\boldsymbol{\hat{z}})}  d \boldsymbol{x} d \boldsymbol{\hat{z}} \\
 = & - \int p(\boldsymbol{y}| \boldsymbol{\hat{z}})p(\boldsymbol{\hat{z}}) \log p(\boldsymbol{y} | \boldsymbol{\hat{z})}  d \boldsymbol{y} d \boldsymbol{\hat{z}} + \beta \int p_{\boldsymbol{\phi}}(\boldsymbol{\hat{z}}|\boldsymbol{x})p(\boldsymbol{x}) \log \frac{p_{\boldsymbol{\phi}}(\boldsymbol{\hat{z}} | \boldsymbol{x)}}{p(\boldsymbol{\hat{z}})}  d \boldsymbol{x} d \boldsymbol{\hat{z}} - H(Y) \\
 = & \underbrace{ - \int p(\boldsymbol{y}| \boldsymbol{\hat{z}})p(\boldsymbol{\hat{z}}) \log q_{\theta}(\boldsymbol{y} | \boldsymbol{\hat{z})}  d \boldsymbol{y} d \boldsymbol{\hat{z}} + \beta \int p_{\boldsymbol{\phi}}(\boldsymbol{\hat{z}}|\boldsymbol{x})p(\boldsymbol{x}) \log \frac{p_{\boldsymbol{\phi}}(\boldsymbol{\hat{z}} | \boldsymbol{x)}}{q(\boldsymbol{\hat{z}})}  d \boldsymbol{x} d \boldsymbol{\hat{z}} }_{\mathcal{L}_{VIB}(\boldsymbol{\phi},\boldsymbol{\theta})} \\
 & - \underbrace{ \int p(\boldsymbol{y}| \boldsymbol{\hat{z}})p(\boldsymbol{\hat{z}}) \log \frac{p(\boldsymbol{y} | \boldsymbol{\hat{z})}}{q_{\theta}(\boldsymbol{y} | \boldsymbol{\hat{z})}}  d \boldsymbol{y} d \boldsymbol{\hat{z}}}_{D_{KL}(p(\boldsymbol{y}|\boldsymbol{\hat{z}})\|q_{\boldsymbol{\theta}}(\boldsymbol{y}|\boldsymbol{\hat{z}})) \geq 0} - \beta \underbrace{ \int p_{\boldsymbol{\phi}}(\boldsymbol{\hat{z}}|\boldsymbol{x})p(\boldsymbol{x}) \log \frac{p(\boldsymbol{\hat{z}})}{q(\boldsymbol{\hat{z}})}  d \boldsymbol{x} d \boldsymbol{\hat{z}}}_{D_{KL}(p(\boldsymbol{\hat{z}})\|q(\boldsymbol{\hat{z}})) \geq 0}  - \underbrace{H(Y)}_{\text{constant}}
 .
\end{aligned}
\end{equation}
\rule{\textwidth}{0.5pt}
\end{minipage}
\end{table*}

\section{Conclusions}
\label{conclusion}

In this work, we investigated task-oriented communication for edge inference, where a low-end edge device transmits the extracted feature vector of a local data sample to a powerful edge server for processing.
Our proposed methodology is built upon the information bottleneck (IB) framework, which provides a principled way to characterize and optimize a new rate-distortion tradeoff in edge inference. Assisted by a variational approximation with a log-normal distribution as the variational prior to promote sparsity in the output feature, we obtained a tractable formulation that is amenable to end-to-end training, named variational feature encoding. We further extended our method to develop a variable-length variational feature encoding scheme based on the dynamic neural networks, which makes it adaptive to dynamic channel conditions. The effectiveness of our methods was verified by extensive simulations on image classification datasets.

Through this study, we would like to advocate for rethinking the communication system design for emerging applications such as edge inference. In these applications, communication will keep playing a critical role, but it will serve for the downstream task rather than for data reconstruction as in the classical communication setting. Thus we should take a task-oriented perspective to design the communication module for such applications. New design tools and methodologies will be needed, and the IB framework is a promising candidate. It bridges machine learning and information theory, and leverages theory and tools from both fields.
There are many interesting future research directions on this exciting topic, e.g., to apply the IB-based framework to the scenario with multiple devices, to develop a theoretical understanding of the new rate-distortion tradeoff, to improve the robustness of the method, etc.

\appendices
\section{Derivation of the Variational Upper Bound}
\label{appendix a}

Recall that the IB objective in (\ref{IB_1}) has the form $\mathcal{L}_{IB}(\boldsymbol{\phi}) = -I(\hat{Z},Y)+\beta I(\hat{Z},X)$. Writing it out in full, the derivation is shown in (\ref{appendix:vib}). 
$\mathcal{L}_{VIB}(\boldsymbol{\phi,\theta})$ in this formulation is the VIB objective function in (\ref{loss_in_3}).
As the KL-divergence is nonnegative and the entropy of $Y$ is a constant, $\mathcal{L}_{VIB}(\boldsymbol{\phi,\theta})$ is a variational upper bound of the IB objective $\mathcal{L}_{I B}(\boldsymbol{\phi})$.

\section{MLP Structure of the Function $\bm{g\big(}\sigma^{2}\bm{\big)}$}
\label{appendix b}
We parameterize $\bm{g\big(}\sigma^{2}\bm{\big)}$ by a $K$-layer MLP, and thus it can be written as a composition of $K$ non-linear functions:
\begin{equation*}
     \bm{g\big(}\sigma^{2}\bm{\big)} = \bm{h_{K}} \circ \bm{h_{K-1}} \cdots \bm{h_{1}\big(}\sigma^{2}\bm{\big)},
\end{equation*}
where $\bm{h_{k}}$ represents the $k$-th layer in the MLP and has $\bm{h_{k}\left({x}\right)} =\operatorname{tanh}\left(\boldsymbol{W^{(k)}} \boldsymbol{x}\right)${\footnote{$\tanh(x)=\frac{e^{x}-e^{-x}}{e^{x}+e^{-x}}$ and $\tanh^{\prime}(x)=1 - \tanh(x)$. For simplicity, we define $\tanh(x)$ and $\tanh^{\prime}(x)$ as element-wise functions.}}.
To maintain the desired properties of the proposed VL-VFE method, each function $g_{j}(\sigma^{2})$ (the $j$-th output dimension of the vector function $\bm{g\big(}\sigma^{2}\bm{\big)}$ should be non-negative and increase with the noise variance $\sigma^2$. Therefore, functions $g_{j}(\sigma^{2})$ should satisfy the following constraints:
\begin{equation*}
    g_{j}(\sigma^{2}) \geq 0; \quad g_{j}^{\prime}(\sigma^{2}) = \frac{\partial g_{j}(\sigma^{2})}{\partial \sigma^{2}} \geq 0.
\end{equation*}
The function ${g(\sigma_{j}^{2})}$ can be writtern as follows:
\begin{equation*}
    \quad g_{j}(\sigma^{2}) = \bm{h_{K,j}} \circ \bm{h_{K-1}} \cdots \bm{h_{1}\big(}\sigma^{2}\bm{\big)},
\end{equation*}
where $\bm{h_{K,j}}$ is $j$-th output dimension of $\bm{h_{K}}$.
The derivative of $g_{j}(\sigma^{2})$ can be obtained using the chain rule:
\begin{equation*}
    \quad g_{j}^{\prime}(\sigma^{2}) = \bm{h_{K,j}^{\prime}} \circ \bm{h_{K-1}^{\prime}} \cdots \bm{h_{1}^{\prime}\big(}\sigma^{2}\bm{\big)},
\end{equation*}
where we denote the Jacobian matrix of $\bm{h_{k}}$ as $\bm{h_{k}^{\prime}}$, and $\bm{h_{K,j}^{\prime}}$ is the $j$-th row of $\bm{h_{K}^{\prime}}$.
The derivatives work out as follows:
\begin{equation*}
\begin{aligned}
\bm{h_{k}^{\prime}({x}) } &=\operatorname{diag} \left(\operatorname{tanh}^{\prime} \left(\boldsymbol{W^{(k)}} \boldsymbol{x}\right) \right) \cdot \boldsymbol{W^{(k)}}.
\end{aligned}
\end{equation*}
To guarantee that each $g_{j}(\sigma^{2})$ is a non-negative increasing function, we set $\bm{W^{(k)}} = \operatorname{abs}(\bm{\widehat{W}^{(k)}})$, which means that $g_{j}\left(\sigma^{2}\right)$ outputs a non-negative value, and all entries in Jacobian matrices are non-negative\footnote{$\operatorname{abs}(\cdot)$ denotes the element-wise absolute function. $\boldsymbol{\widehat{W}^{(k)}}$ are the actual parameters in the $K$-layer MLP.}.

\begin{figure*}[t]
\centering
\begin{minipage}[t]{0.4\textwidth}
\centering
\includegraphics[width=1\linewidth]{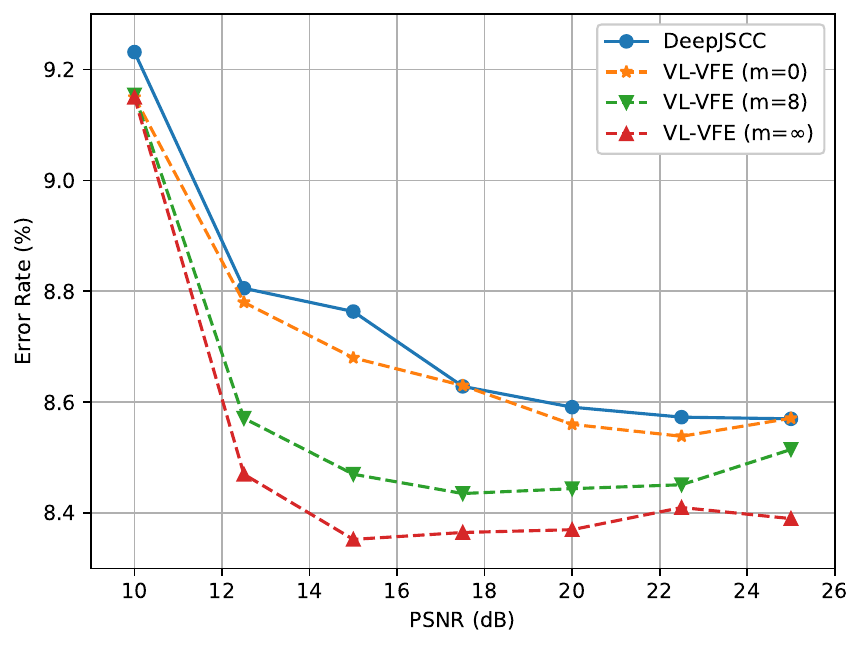}
\caption{Error rate as a function of the channel PSNR in dynamic channel conditions.}
\label{robust_acc}
\end{minipage}
\hspace{.1in}
\begin{minipage}[t]{0.4\textwidth}
\centering
\includegraphics[width=1\linewidth]{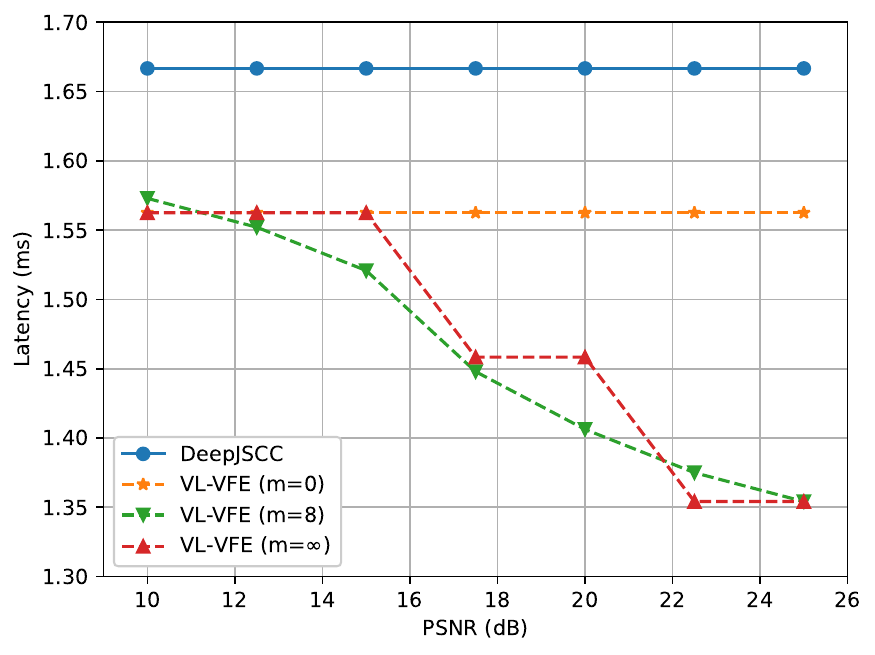}
\caption{Communication latency as a function of the channel PSNR in dynamic channel conditions.}
\label{robust_latency}
\end{minipage}
\end{figure*}

\begin{figure*}[t] 
\centering
\subfloat[The rate-distortion curves with PSNR = 20 dB.]{
\centering
\includegraphics[width=0.382\linewidth]{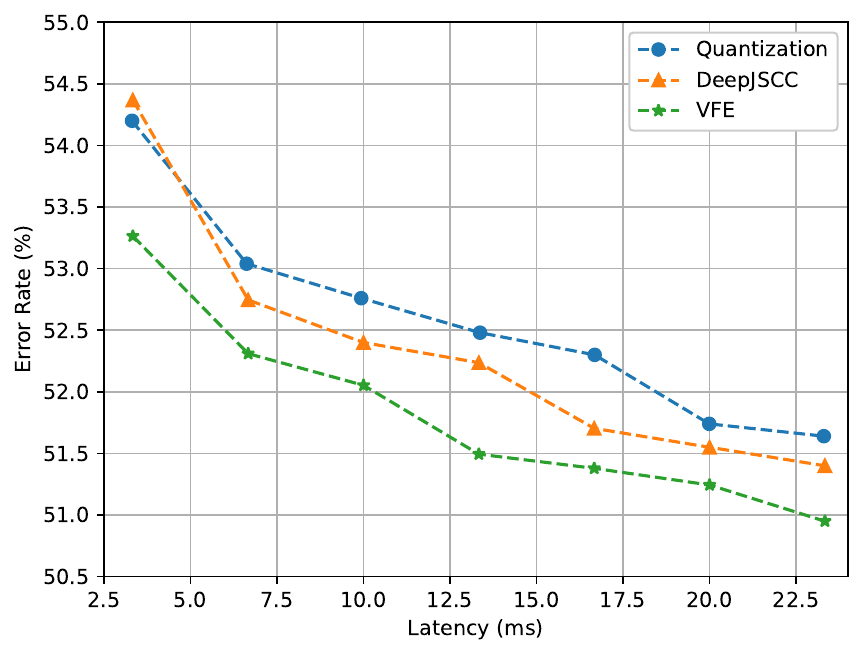}
\label{tiny_imagenet_exp_static}
}
\hspace{.1in}
\subfloat[Communication latency and error rate as a function of the channel PSNR in dynamic channel conditions.]{
\centering
\includegraphics[width=0.4\linewidth]{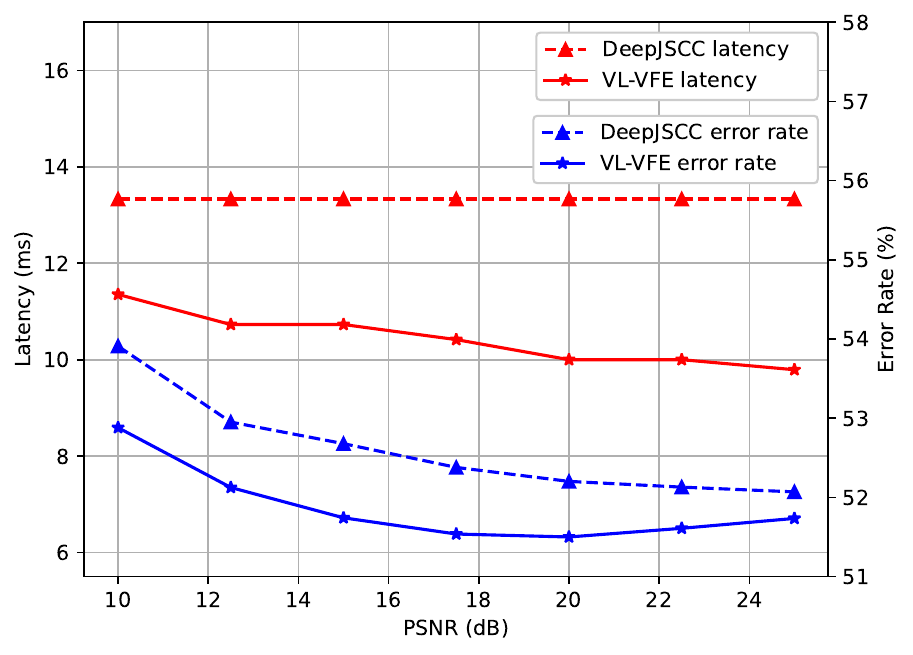}
\label{tiny_imagenet_exp_dynamic}
}
\caption{Experimental results of the classification task on the Tiny ImageNet dataset.}
\end{figure*}

\section{Robustness of the VL-VFE method given inaccurate channel noise variance}
\label{appendix C}
We conduct the experiments to evaluate the robustness of the proposed method given inaccurate channel noise variance.
In particular, by assuming $m$ pilot symbols are transmitted from the mobile device for noise variance estimation, and adopting the uniformly minimum-variance unbiased estimator, the noise variance is estimated as $\hat{\sigma}^{2} = \frac{1}{m-1} \sum_{i}^{m} (\hat{z}_{i,p} - {z}_{i,p})^{2}$, where $\hat{z}_{i,p}$ and ${z}_{i,p}$ correspond to the $i$-th transmitted and received pilot symbols, respectively. It can be easily verified that $\mathbf{E}\left[\hat{\sigma}^{2}\right] = \sigma^{2}$ and $p(\hat{\sigma}^{2}|{\sigma}^{2}) = \frac{\sigma^{2}}{m-1}\chi^{2}(m)$, where $\chi^{2}(m)$ denotes the chi-square distribution with $m$ degrees of freedom.
The variance of $\hat{\sigma}^{2}$ reduces as $m$ increases, i.e., the noise variance estimation becomes more accurate.
With the inaccurate noise variance $\hat{\sigma}^{2}$ at the transmitter, we test the performance of the proposed VL-VFE method based on the CIFAR-10 image classification task for the following three cases:
\begin{itemize}
\item VL-VFE ($m=0$): This corresponds to the case that the transmitter has no knowledge about the noise variance, and the PSNR is set to be 10 dB for feature encoding;
\item VL-VFE ($m=8$): The noise variance is estimated via 8 pilot symbols, which corresponds to the case of imperfect channel knowledge for feature encoding;
\item VL-VFE ($m=\infty$): This corresponds to the case of perfect channel knowledge for feature encoding.
\end{itemize}
Following the experimental settings in Section V, we also adopt DeepJSCC as the baseline in comparison.
The experimental results on the error rate and feature transmission latency are shown in Fig. \ref{robust_acc} and Fig. \ref{robust_latency}, with the new findings summarized as follows:

\begin{itemize}
\item The proposed method achieves lower communication latency compared with DeepJSCC in all the three cases in the dynamic channel conditions;
\item While reducing the number of pilot symbols to 8 incurs performance degradation to the proposed method due to the inaccurate noise variance, the proposed method still achieves a much better rate-distortion tradeoff than DeepJSCC;
\item Even when the transmitter has no knowledge of the noise variance, i.e., $m=0$, the proposed method still shows a comparable performance as DeepJSCC.
\end{itemize}

In conclusion, these experimental results demonstrate that our proposed method is robust against the inaccurate channel knowledge, i.e., the channel noise variance.

\section{Additional Experiments on Tiny ImangeNet dataset}
\label{appendix D}

We further evaluate the performance of the proposed variational feature encoding (VFE) method and variable-length variational feature encoding (VL-VFE) method on the Tiny ImageNet classification task \cite{tiny_imagenet}.
Tiny ImageNet contains 200 image classes, a training dataset of 100,000 images, and a validation dataset of 10,000 images. All images are of size 64 $\times$ 64.
We select the ResNet18 as the backbone for this task, which can achieve the top-1 accuracy of around 50.5\%. 
The whole neural network structure is shown in Table \ref{network_structure_tiny_imagenet}.
Following the basic settings in Section V, we compare our proposed methods with \textbf{DeepJSCC} and \textbf{Learning-based Quantization}.
The initialized feature dimension of the proposed methods is 224 in this set of experiments.
Fig. \ref{tiny_imagenet_exp_static} shows the rate-distortion curves in the static channel condition (PSNR = 20 dB), where our VFE method changes the feature dimension by adjusting the $\beta$ value in the range of $[10^{-4},10^{-3}]$.
Similar to the previous results on the MNIST and CIFAR-10 datasets, our proposed VFE method outperforms the baselines by achieving a better rate-distortion tradeoff.
In the dynamic channel conditions, we set $\beta = 5 \times 10^{-4}$ in the training phase when PSNR is changing from 10 dB to 25 dB. Fig. \ref{tiny_imagenet_exp_dynamic} shows that the proposed  VL-VFE  method achieves higher accuracy and lower latency compared with DeepJSCC.

\begin{table}[]
\selectfont
\begin{center}
\caption{The DNN structure for Tiny ImageNet classification task.}
\resizebox{0.475\textwidth}{!}{
\begin{tabular}{c|c|c}
\hline
                                      & \textbf{Layer}                                  & \begin{tabular}[c]{@{}c@{}}\textbf{Output} \\ \textbf{Dimensions}\end{tabular} \\ \hline
\multirow{2}{*}{\begin{tabular}[c]{@{}c@{}}\textbf{On-device} \\ \textbf{Network}\end{tabular}}
                                      & [ResNet Building Block] $\times$ 5                  & 512 $\times$ 4 $\times$ 4     \\ \cline{2-3} 
                                      & Pooling + Fully-connected Layer + Tanh & n                 \\ \hline
\multirow{2}{*}{\begin{tabular}[c]{@{}c@{}}\textbf{Server-based} \\ \textbf{Network}\end{tabular}}
                                      & Fully-connected Layer + ReLU                          & 512               \\ \cline{2-3} 
                                      & Fully-connected Layer + Softmax        & 200                \\ \hline
\end{tabular}
}
\label{network_structure_tiny_imagenet}
\end{center}
\end{table}


\ifCLASSOPTIONcaptionsoff
  \newpage
\fi



%

\bibliographystyle{./bibtex/IEEEtran}
\bibliography{./bibtex/IEEEabrv,ref}

%




\end{document}